\documentclass[aps, pre, twocolumn, floatfix, groupedaddress]{revtex4-2}

\usepackage[bottom]{footmisc}
\usepackage{lipsum}
\usepackage{float}
\usepackage{graphicx}
\usepackage{algorithmic}
\usepackage{amssymb, amsmath, amsfonts}
\usepackage[caption=false]{subfig} 
\usepackage{pgfplots} 
\usepackage{xcolor}
\usepackage{mathtools}
\usepackage{bm}
\usepackage{amsopn}
\usepackage[normalem]{ulem} 
\newcommand{\Mod}[1]{\ (\mathrm{mod}\ #1)}

\DeclareMathAlphabet{\mathdutchcal}{U}{dutchcal}{m}{n} \SetMathAlphabet{\mathdutchcal}{bold}{U}{dutchcal}{b}{n}
\DeclareMathAlphabet{\mathdutchbcal}{U}{dutchcal}{b}{n} 

\makeatletter
\renewcommand*\env@matrix[1][\arraystretch]{%
  \edef\arraystretch{#1}%
  \hskip -\arraycolsep
  \let\@ifnextchar\new@ifnextchar
  \array{*\c@MaxMatrixCols c}}
\makeatother

\begin{document}


\title{Exact Spatio-Temporal Dynamics of Lattice Random Walks in Hexagonal and Honeycomb Domains}


\author{Daniel Marris}
\author{Seeralan Sarvaharman}
\affiliation{Department of Engineering Mathematics, University of Bristol, BS8 1UB}
\author{Luca Giuggioli}
\affiliation{Bristol Centre for Complexity Sciences and Department of Engineering Mathematics, University of Bristol,
BS8 1UB}


\date{\today}

\begin{abstract}
    A variety of transport processes in natural and man-made systems are intrinsically random. To model their
    stochasticity, lattice random walks have been employed for a long time, mainly by considering Cartesian lattices.
    However, in many applications in bounded space the
    geometry of the domain may have profound effects on the dynamics and ought to be
    accounted for. We consider here the cases of the six-neighbour (hexagonal) and three-neighbour (honeycomb) lattice,
    which are utilised in models ranging from adatoms diffusing in metals and excitations diffusing on single-walled
    carbon nanotubes to animal foraging strategy and the formation of territories in scent-marking organisms. In these
    and other examples, the main theoretical tool to study the dynamics of lattice random walks in hexagonal geometries
    has been via simulations. Analytic representations have in most cases been inaccessible, in particular in bounded
    hexagons, given the complicated `zig-zag' boundary conditions that a walker is subject to. Here we generalise the
    method of images to hexagonal geometries and obtain closed-form expressions for the occupation probability, the
    so-called propagator, for lattice random walks both on hexagonal and honeycomb lattices with periodic, reflective
    and absorbing boundary conditions. In the periodic case, we identify two possible choices of image placement and
    their corresponding propagators. Using them, we construct the exact propagators for the other boundary conditions,
    and we derive transport related statistical quantities such as first passage probabilities to one or multiple
    targets and their means, elucidating the effect of the boundary condition on transport properties.
\end{abstract}

\maketitle

\section{Introduction}\label{sec: Intro} 
Popularised in the 1920s by P{\'o}lya \cite{polya1921aufgabe}, lattice random walks (LRW) are widely used in the
mathematics \cite{feller1967introduction} and physics \cite{rws_on_latticesII, barry_hughes_book, weiss1994aspects}
literature. Owing to their versatility as a special class of Markov chains, one finds applications of LRW across a
multitude of disciplines including animal ecology \cite{okubo2001diffusion, giuggioli2014consequences}, cell biology
\cite{wu2015statistical}, actuarial science \cite{avram2019first} and social sciences \cite{newman2005measure}. While
analytic representations of the dynamics of LRW has been studied for a long time, recent advances in the exact
description of their dynamics in finite $d$-dimensional hypercubic lattices \cite{sarvaharman2020closed, LucaPRX,
das2022discrete, das2023dynamics} have brought renewed interest. Following these advances, many computationally
challenging problems such as transmission and encounter dynamics between LRW pairs \cite{luca_transmission} and the
dynamics of interactions with inert spatial heterogeneities can now be tackled analytically \cite{sree_reflect}. 

These developments are, however, limited to Cartesian lattices with little attention given to other geometries. In $d=2$
dimensions, two such important cases are the hexagonal and the honeycomb lattices, often found to be used
interchangeably in the literature \cite{batchelor2002exact, honeycomb_unbounded_paper, rws_on_latticesIII}. To avoid any
confusion, in the present work we refer to a hexagonal lattice when each site has six nearest-neigbours and to a
honeycomb lattice when each site has three nearest-neighbours.

Random walks on both lattices have been employed to study many stochastic processes. The hexagonal lattice has been used
to represent adatom diffusion in metals \cite{zaluska1999collective}, space usage and foraging of animals
\cite{PRASAD2006241}, territory formation in scent-marking organisms \cite{robles2018phase, kenkre2021theory}, and
substrate diffusion across artificial tissue \cite{astor2000developmental}, while the honeycomb LRW has been utilised for
particle movement in ice and graphite \cite{bercu2021asymptotic} and the diffusion of excitons on a single-walled carbon
nanotube (SWCN) \cite{cotfas2000random, cotfas2005alternate}. Both lattices are also of interest in the context of
self-avoiding walks \cite{guttmann1984two, duminil2012connective}. 

For both lattices, when unbounded, the walk statistics have been studied by mapping the
dynamics onto a square lattice. To model a hexagonal walk, two of the eight permissible movement directions in a next
nearest-neighbour walk are removed \cite{rws_on_latticesIII, guttmann2010lattice}, while a brick-like structure of
positive and negative sites is created to represent the honeycomb LRW \cite{barry_hughes_book, de1986self}. Another
approach, applicable to the honeycomb lattice, models the domain as a bipartite network allowing the construction of the
propagator generating function for an unbounded walker that always moves \cite{honeycomb_unbounded_paper}. 

For many of the applications stated earlier, the movement statistics are heavily affected by the size of the underlying
spatial domain. In the literature, analytical attempts to take into consideration the finiteness of the space have been
rare, largely due to the non-orthogonal configuration of lattice sites, which leads to complex `zig-zag' boundaries. One
example, which aims to account for the dynamics at the boundary, considers a four walled domain with two opposing walls
made up of absorbing sites and the other two walls representing a periodic domain. Via the use of a technique to solve
inhomogenous linear partial difference equations \cite{mccrea1940xxii}, analytic expressions for the expectation value
that the walk reaches a lattice site before getting absorbed have been obtained \cite{henry2003random}.

To represent faithfully the finiteness of the hexagonal space, we introduce here a framework for the analytic representation of the
spatio-temporal dynamics of LRW in hexagonal geometries with true `zig-zag' boundaries. We utilise a non-orthogonal
hexagonal coordinate system \cite{her1995geometric, coordinate_system2} for both lattices and model the honeycomb
lattice via the inclusion of internal states \cite{weiss1983random, barry_hughes_book}. By deriving an extension of the
method of images \cite{montroll1979enriched} to hexagonal space we find closed-form expressions for the propagator, in
periodically bounded random walks. We generalise the defect technique \cite{kenkre2021memory, luca_transmission,
sree_reflect} to hexagonal space and random walks with internal states and obtain analytically the propagator generating
function for both LRW with absorbing and reflecting boundary conditions. Various transport properties in both lattices
are also analysed. 

The outline of the paper is as follows. In Sec. \ref{sec: Hex_Geometry} we introduce the coordinate system used to
parameterise the lattices. Section \ref{sec: Unbounded} is devoted to the analysis of the unbounded lattice Green's
function or propagator for both the hexagonal and honeycomb random walk. In Sec. \ref{sec: Periodic}, closed-form
expressions for the propagator, in two representations of periodically bounded domains are obtained. In Secs. \ref{sec:
Absorb} and \ref{sec: Reflect}, we derive the propagators for absorbing and reflecting domains, respectively. Transport
properties are studied in Sec. \ref{sec: First_Passage} where we present the analytic representation of the first
passage probability, or first-hitting time, to a target, while in Sec. \ref{sec: MFPT} we derive closed-from expressions
for the mean first passage time to one target and employ it to study the mean first passage to multiple targets.

\section{Hexagonal Coordinate System}\label{sec: Hex_Geometry}  
\begin{figure*}
	\includegraphics[width = 0.8\textwidth]{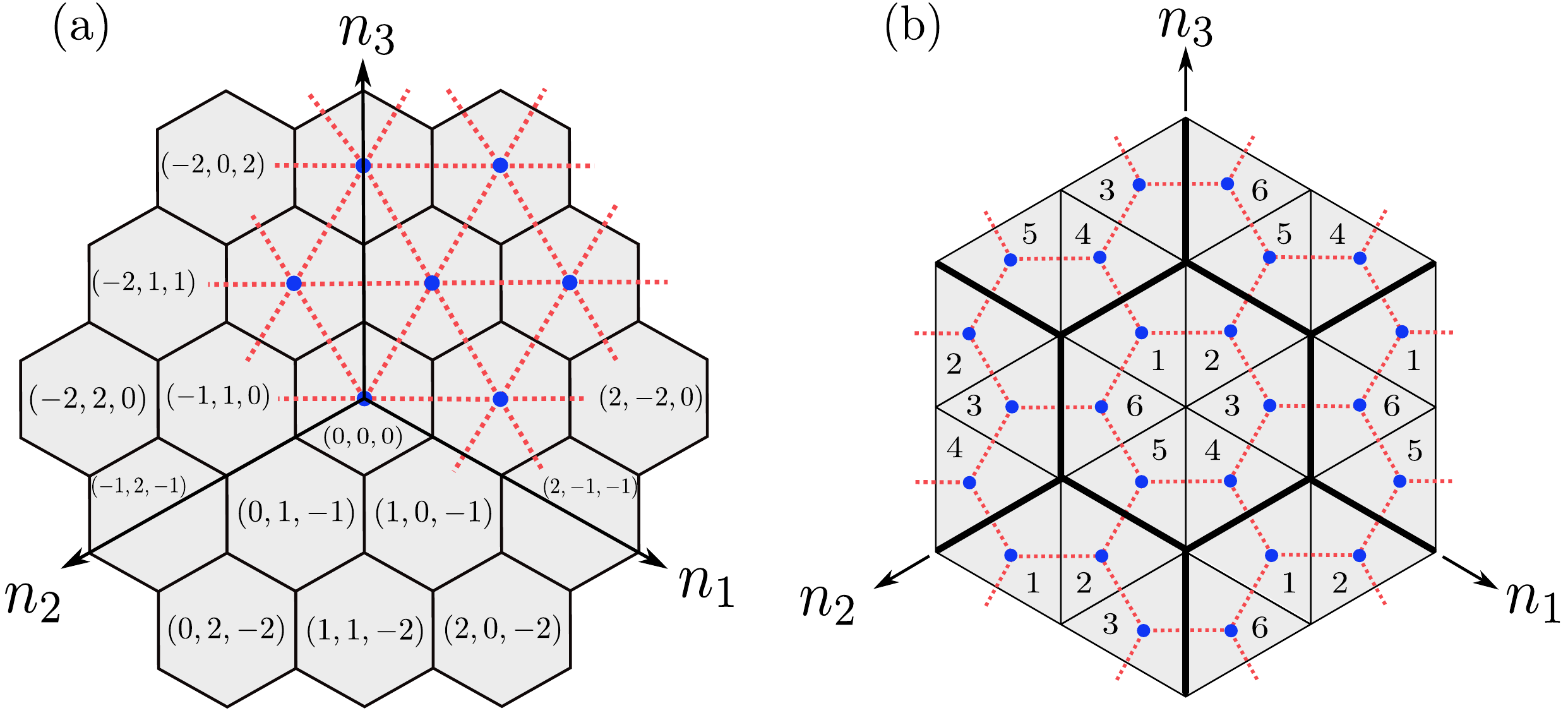}
	\caption{(Colour online). A schematic representation of the hexagonal lattice, panel (a), and the honeycomb lattice,
	panel (b). In (a), we show the HCC in $\mathbb{R}^2$ with three non-orthogonal axes. Coordinate labels are shown on
	some lattice sites, while permissible movement directions for the hexagonal lattice are shown on others with dotted
	lines. For clarity, we omit arrows depicting the option of staying on lattice sites. In (b) we show the honeycomb
	lattice. Here we show the $(0, 0, 0)$ Wigner-Seitz cell, half of its six neighbours and their labelled internal
	states. Permissible movement directions are again shown through dotted lines and similarly to panel (a), arrows
	representing the option of remaining on any site are removed. The boundaries of the hexagonal Wigner-Seitz cell are
	shown in bolder lines.}
	\label{fig: hex_honey_coords}
\end{figure*}

A convenient coordinate system for hexagonal lattices, designed for application in computer
graphics~\cite{her1995geometric, coordinate_system2}, is given by three linearly dependent integer coordinates $(n_1,
n_2, n_3)$ such that $n_1 + n_2 + n_3 = 0$. One can represent these coordinates on two different axis sets: an oblique
plane in $\mathbb{R}^3$ or three axes lying 60 degrees apart in $\mathbb{R}^2$. We take the latter, whose visual
representation can be found in Fig. \ref{fig: hex_honey_coords}(a), and we refer to the coordinate system as hexagonal
cubic coordinates (HCC). The HCC system is related to $\mathbb{R}^2$ Cartesian coordinates $(x_1, x_2)$ via the
transformation 
\cite{cube_hex_coords} 
\begin{equation}
\label{eq: HCC_to_cart}
n_1 = -\frac{x_2}{2} + \frac{\sqrt{3}x_1}{2}; \; \: \; n_2 = x_2; \; \: \; n_3 = -\frac{x_2}{2} - \frac{\sqrt{3}x_1}{2},
\end{equation}
where $n_1, n_2, n_3 \in \mathbb{Z}$ and $x_1, x_2 \in \mathbb{R}$, which enables convenient plotting.
\subsection{Finite Hexagonal Lattice} \label{sec: hexagonal_lattice} 
We model permissible jumps to neighbouring sites taking place between the centroid of the so-called Wigner-Seitz cell
(WS) and its six neighbours, shown in Fig. \ref{fig: hex_honey_coords}(a), giving a coordination number $Z=6$.
We also allow for the option of remaining on the lattice site at each timestep. The size of the
finite domain is controlled by the single parameter $R$, the circumradius of the hexagon. The corners lie at $(n_1, n_2,
n_3) = (\pm R, 0, \mp R)$, $(n_1, n_2, n_3) = (\pm R, \mp R, 0)$, and $(n_1, n_2, n_3) = (0, \pm R, \mp
R)$. 

\subsection{Finite Honeycomb Lattice}\label{sec: honey_lattice} 
The honeycomb lattice has a coordination number $Z=3$ and each lattice site can be thought of as a vertex of the
hexagonal WS cell, making the honeycomb lattice the dual of the hexagonal lattice \cite{nagy2015cellular}. We model this
lattice as a tessellation of even ($\rhd $) and odd ($\lhd$) triangles by utilising the HCC and including internal
states. To avoid confusion, we refer to the hexagonal WS cell, with coordinates $(n_1, n_2, n_3)$ as a lattice site,
which contains six internal states labelled $m_i$, $i = \{1, ...6\}$ starting from the top left triangle and going
clockwise round the unit cell, shown in Fig. \ref{fig: hex_honey_coords}(b). At each timestep the walker has four
permissible actions: to remain on the same lattice site and state, to move to either of two adjacent states in the same
site or to move to an adjacent state in an adjacent site. The number of locations the walker can reach in the honeycomb
lattice is six times bigger than in the hexagonal lattice with the same circumradius. 
\section{Dynamics in Unbounded Space}\label{sec: Unbounded} With later sections exploiting the analytic representation
of the occupation probability for the infinite lattice to construct bounded propagators, we show here the procedure to
obtain the unbounded case.
\subsection{Hexagonal Lattice}
The Master equation governing the evolution of the site occupation probability, $Q(n_1, n_2, n_3, t)$, for the unbounded
hexagonal lattice is represented by
\begin{widetext}
    \begin{equation}
        \label{eq: original-master-eq}
            \begin{aligned}
                Q(&n_1, n_2, n_3, t+1) = \frac{q}{6}\bigg[Q(n_1 - 1, n_2 , n_3+ 1, t) +  Q(n_1 , n_2 - 1, n_3+ 1, t) + Q(n_1 + 1, n_2- 1, n_3 , t)\\
                            & +  Q(n_1 + 1, n_2, n_3 - 1, t) +  Q(n_1, n_2 + 1, n_3 - 1, t) + Q(n_1 - 1, n_2 + 1, n_3, t)\bigg] +
                             (1-q) Q(n_1, n_2, n_3, t),
            \end{aligned}
        \end{equation}
\end{widetext}
where $q$ ($0< q \leq 1$), determines the probability of movement, that is $q=1$ represents a walker changing lattice
site at every timestep. Eq. (\ref{eq: original-master-eq}) is subject to the initial condition $Q(n_1, n_2, n_3, 0) =
\delta_{n_{_1} n_{0_1}}\delta_{n_{_2} n_{0_2}}\delta_{n_{_3} n_{0_3}}$, where $\delta_{i j}$ is the Kronecker delta and
$n_{0_1} + n_{0_2} + n_{0_3} = 0$. 

While Eq. (\ref{eq: original-master-eq}) is well suited for the infinite lattice, in the bounded cases the linear
relationship between the coordinates, $n_3 = -n_1 - n_2$, makes it necessary (see Appendix \ref{app: periodic}) to drop
the $n_3$ dependence and re-write the Master equation with a two coordinate representation given by
\begin{widetext}
    \begin{equation}
        \label{eq: two-variable-master-eq}
            \begin{aligned}
                Q(&n_1, n_2, t+1) = \frac{q}{6}\bigg[Q(n_1 - 1, n_2, t) +  Q(n_1 , n_2 - 1, t) + Q(n_1 + 1, n_2  - 1, t) +  \\ &Q(n_1 + 1, n_2, t) + Q(n_1, n_2 + 1, t) + Q(n_1 - 1, n_2 + 1, t)\bigg] + (1-q) Q(n_1, n_2, t).
            \end{aligned}
        \end{equation}
\end{widetext}
After applying the discrete Fourier transform $\widehat{f}(k) = \sum_{n=-\infty}^{\infty}e^{-i k n}f(n)$ and the
unilateral $z$-transform $\widetilde{f}(z) = \sum_{t=0}^{\infty}z^t f(t)$, we solve Eq. (\ref{eq:
two-variable-master-eq}) to obtain the hexagonal lattice Green's function as a double integral 
\begin{equation}
    \label{eq: unbounded_prop}
    \widetilde{Q}_{\bm{n}_{0}}(n_1, n_2, z) = \frac{1}{(2\pi)^2}\int_{-\pi}^{\pi} \int_{-\pi}^{\pi} \frac{e^{i[(\bm{n} - \bm{n}_0)\cdot \bm{k}]}}{1-z \mu(k_1, k_2)}dk_1dk_2,
\end{equation}
where  $\bm{k} = (k_1, k_2)^{\intercal}$ and $\bm{n}-\bm{n}_0 = (n_{_1} - n_{0_1}, n_{_2} - n_{0_2})$ and $\mu(k_1, k_2)
= 1-q+\frac{q}{3}\left[\cos(k_1-k_2) + \cos(k_1) + \cos(k_2)\right]$ is the so-called structure function, or discrete
Fourier transform of the individual step probabilities \cite{rws_on_latticesII}. Equation (\ref{eq: unbounded_prop})
reduces to known results when $q=1$ \cite{rws_on_latticesIII}. 
\subsection{Honeycomb Lattice}
The general form of the Master equation for a LRW with internal states has the vectorial form \cite{internal_states2,
barry_hughes_book} 
\begin{equation}
\label{eq: bm_is_master}
	\bm{ \mathcal{Q}}(\bm{n}, t+1) = \sum_{\bm{n}'}\mathbb{W}(\bm{n}, \bm{n}')\bm{ \mathcal{Q}(\bm{n}'}, t),
\end{equation}
where $\mathbb{W}(\bm{n}, \bm{n}^{\prime})$  represents all possible movement from $\bm{n}^{\prime}$ to site $\bm{n}$ at
each moment in time and $\bm{ \mathcal{Q}(\bm{n}'}, t)$ is a column vector representing the occupation probability of
each internal state in site $\bm{n}'$ at time $t$. For the honeycomb lattice, Eq. (\ref{eq: bm_is_master}) reduces to
\begin{widetext}
    \begin{equation}
        \label{eq: honey-master-kron}
            \begin{aligned}
                \bm{\mathcal{Q}}(&n_1, n_2, t+1) = \frac{q}{3}\bigg[\mathbb{A}_{1,4}\cdot\bm{\mathcal{Q}}(n_1 - 1, n_2, t) +  \mathbb{A}_{2,5}\cdot\bm{\mathcal{Q}}(n_1 , n_2 - 1, t) +  \mathbb{A}_{3,6}\cdot\bm{\mathcal{Q}}(n_1 + 1, n_2  - 1, t) + \\ & \mathbb{A}_{4,1}\cdot\bm{\mathcal{Q}}(n_1 + 1, n_2, t) +\mathbb{A}_{5,2}\cdot\bm{\mathcal{Q}}(n_1, n_2 + 1, t) + \mathbb{A}_{6,3}\cdot\bm{\mathcal{Q}}(n_1 - 1, n_2 + 1, t)\bigg] + \mathbb{B}\cdot\bm{\mathcal{Q}}(n_1, n_2, t),
            \end{aligned}
        \end{equation}
\end{widetext}
where $\mathbb{A}_{i,j}$ is a $6\times6$ matrix with value one at index $i,j$ that represents the movement from state
$i$ in one WS cell to state $j$ in a new WS cell, as depicted in Fig. \ref{fig: hex_honey_coords}(b). $\mathbb{B}$,
which represents the movement within one WS cell, is a tridiagonal Toeplitz matrix with perturbed corners where
$\mathbb{B}_{i,i} = 1-q$ with $i \leq 1 \leq 6$, $\mathbb{B}_{i+1,i} = \mathbb{B}_{i,i+1} = \frac{q}{3}$ with $1 \leq i
\leq 5$, $\mathbb{B}_{1, 6} = \mathbb{B}_{6, 1} = \frac{q}{3}$ and zero elsewhere.

Taking the localised initial condition $\bm{\mathcal{Q}}(n_1, n_2, t=0) = \delta_{\bm{n} \bm{n}_{0}}\bm{U}_{m_0}$, where
$\bm{U}_{m_0}$ is a $6\times 1$ column vector with element $m_0$ exactly one and the rest exactly zero, and following
standard techniques for random walks with internal states (see e.g. \cite{barry_hughes_book}), we obtain the generating
function of the unbounded propagator
\begin{widetext}
    \begin{equation}
\begin{aligned}
\label{eq: honeycomb_unbounded}
	\widetilde{\bm{\mathcal{Q}}}_{\bm{n}_{0}, m_0}(n_1, n_2, z) = \frac{1}{(2\pi)^2}\int_{-\pi}^{\pi}\int_{-\pi}^{\pi} e^{i[(\bm{n} - \bm{n}_0)\cdot \bm{k}]} 
    \left[\mathbb{I} -  z\bm{\mu}(k_1, k_2)\right]^{-1} \cdot \bm{U}_{m_0} dk_1 dk_2,
	\end{aligned}
\end{equation}
\end{widetext}
where $\mathbb{I}$ is the $6\times 6$ identity matrix and the structure function

\begin{equation}
    \begin{aligned}
        &\bm{\mu}(k_1, k_2) = \\&
        \begin{bmatrix}
            1-q& \frac{q}{3}&0 &\frac{q}{3}e^{-i k_1} &0 &\frac{q}{3} \\
            \frac{q}{3} & 1-q & \frac{q}{3} & 0 & \frac{q}{3}e^{-i k_2} & 0 \\
            0 & \frac{q}{3} & 1-q & \frac{q}{3} & 0 & \frac{q}{3}e^{i(k_1 - k_2)} \\
            \frac{q}{3}e^{i k_1} & 0 & \frac{q}{3} & 1-q & \frac{q}{3} & 0 \\
            0 & \frac{q}{3}e^{i k_2} & 0 & \frac{q}{3} & 1-q & \frac{q}{3} \\
            \frac{q}{3} & 0 & \frac{q}{3}e^{-i(k_1 - k_2)} & 0 & \frac{q}{3} & 1-q \\
        \end{bmatrix}.
    \end{aligned}
\end{equation}

\sloppy To access the probability at a unique state $m$ one simply takes the scalar dot product
$\widetilde{\mathcal{Q}}_{\bm{n}_{0}, m_0}(n_1, n_2, m, z) =
\bm{U}_m^{\intercal}\cdot\widetilde{\bm{\mathcal{Q}}}_{\bm{n}_{0}, m_0}(n_1, n_2, z)$.
\section{Periodic Boundary Conditions}\label{sec: Periodic} To impose periodic boundary conditions, we generalise the
method of images \cite{montroll1979enriched} to hexagonal domains. The technique represents a convenient way to impose
boundary conditions on Green's functions. To implement the technique for fully bounded domains, one considers an
infinite set of initial conditions, tessellated across the space, which act concurrently by mirroring the walker's
movement. To tessellate hexagonal lattices in 2-dimensional space, there are two choices for the placement of the
neighbouring domains due to the `zig-zag' nature of the boundaries, which differ depending on whether the images are
located with a so-called left or right shift in relation to one of the axes. To illustrate, let us consider the hexagon
directly above the chosen finite domain. If the bottom right corner of the neighbouring domain is to the right of the
top right corner of the modelled domain, it is referred to as the right shift tessellation, otherwise, it is to the left,
and is referred to as the left shift tessellation (see Appendix \ref{app: image placement} for a pictorial
representation).

Using either tessellation, we construct an infinite number of images of the initial condition and obtain the bounded
periodic propagator
\begin{equation}
\label{eq: periodic_images}
    \widetilde{P}^{(p)}_{\bm{n}_{0}}(n_1, n_2, z) = \sum_{m_1 = -\infty}^{\infty}\sum_{m_2 = -\infty}^{\infty}\widetilde{Q}_{\bm{n}_{0}}(n_1 + \hat{n}_1, n_2 + \hat{n}_2, z) ,
\end{equation}
built by considering the appropriate coordinate transform from a location in the finite hexagonal domain to the
equivalent location in any of the infinite neighbouring domains. For the right shift we find 
\begin{figure*}[ht] 
    \includegraphics[width = 0.8\textwidth]{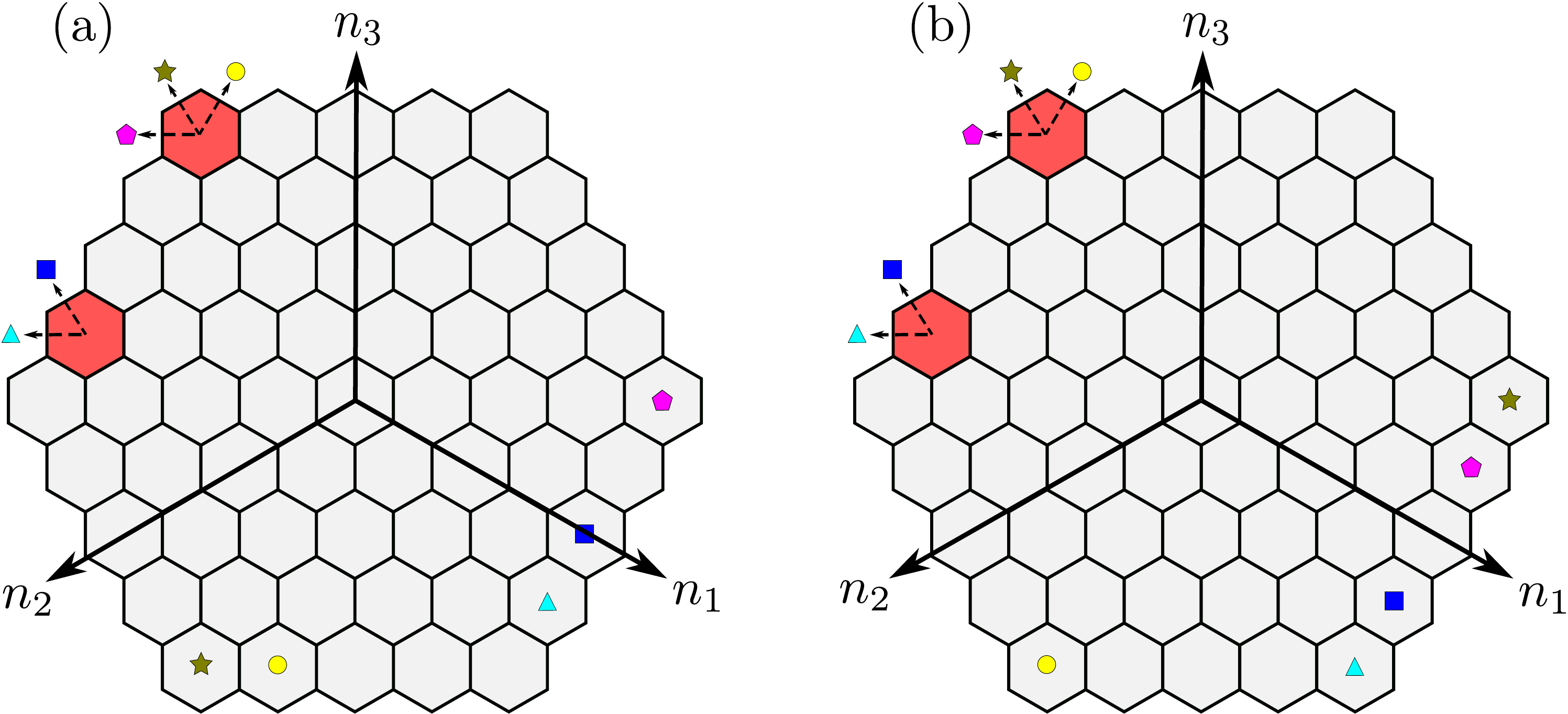}
    \caption{(Colour online). A schematic showing the difference in boundary dynamics for two specific lattice sites between the left shift, panel (a), and the right shift, panel (b). The dashed arrows show permissible directions of movement at the two chosen boundary sites and the shape at the end of the arrow corresponds with the shape indicating where this direction of travel would lead to.}
    \label{fig: periodic_boundary_dynamics}
\end{figure*} 
\begin{equation}
\label{eq: right_shift1}
    \begin{bmatrix}
        \widehat{n}_1 \\ \widehat{n}_2 
    \end{bmatrix} = \begin{bmatrix}
        -Rm_1 +(2R+1)m_2 \\ -(R+1)m_1 - Rm_2
    \end{bmatrix},
\end{equation}
and for the left shift
\begin{equation}
\label{eq: left_shift1}
    \begin{bmatrix}
        \widehat{n}_1 \\ \widehat{n}_2 
    \end{bmatrix} = \begin{bmatrix}
        (2R+1)m_1 -R m_2 \\  -(R+1)m_1 + (2R+1)m_2
    \end{bmatrix},
\end{equation}
where $m_1, m_2 \in \mathbb{Z}$. 

\subsection{Hexagonal Lattice}\label{subsec: hex_periodic} Applying Eq. (\ref{eq: unbounded_prop}) in Eq. (\ref{eq:
periodic_images}) and interchanging the order of integration and summation, as is permissible in generalised function
theory \cite{barry_hughes_book}, one solves (see Appendix \ref{app: periodic}) for the periodic propagator
\begin{widetext}
    \begin{equation}
        \label{eq: periodic-sol-hex-coords}
        \begin{aligned}
             P_{\bm{n}_{0}}^{(p)_{[i]}}(n_1, n_2, & t) = \frac{1}{\Omega} + \frac{2}{\Omega} \sum_{r=0}^{R-1}\sum_{s=0}^{3r+2} \cos\left(\frac{2\pi \left[k_1^{[i]}(r,s)\left(n_1-n_{0_1}\right)+ k_2^{[i]}(r,s)\left(n_2-n_{0_2}\right)\right]}{\Omega}\right) \\
             & \times\left(1-q+\frac{q}{3}\left[\cos\left(\frac{2\pi\left(k_1^{[i]}(r,s)-k_2^{[i]}(r,s)\right)}{\Omega}\right) + \cos\left(\frac{2\pi k_1^{[i]}(r,s)}{\Omega}\right) + \cos\left(\frac{2\pi k_2^{[i]}(r,s)}{\Omega}\right)\right]\right)^t,
        \end{aligned}
        \end{equation}
\end{widetext}   
with $i \in \{\rho, \lambda\}$ indicating the right and left shift respectively. $\Omega$ is the number of lattice
sites in the finite domain, namely $\Omega = 3R^2 + 3R + 1$, and 
\begin{equation}
\label{eq: ks}
\begin{aligned}
	k^{[\rho]}_1(r,s) &= k^{[\lambda]}_2(r,s) = R(s+1) + s - r, \\
    k^{[\rho]}_2(r,s) &= k^{[\lambda]}_1(r,s) = R(2-s+3r) + r + 1 .
\end{aligned}
\end{equation}
We note here that as $k_1(r,s)$ and $k_2(r,s)$ are interchanged under the transition between $\rho$ and $\lambda$, the
structure function is not dependent on this choice. We further note that one can also find $\Omega$
using the so-called centered hexagonal number \cite{teo1985magic,robles2018phase}.

While the dynamics in the bulk are identical, the walker acts differently at the boundaries depending on the chosen
shift (see Fig. \ref{fig: periodic_boundary_dynamics}). As we will see in Secs. \ref{sec: First_Passage} and \ref{sec:
MFPT}, it may lead to marked differences in the first passage probability and mean first passage (MFPT) time.

We note that Eq. (\ref{eq: periodic-sol-hex-coords}) is still valid for the trivial $R= 0$ case where the domain reduces
to a single point at the origin. Here, the two summations disappear and the solution reduces to
$P_{\bm{n}_{0}}^{(p)_{[i]}}(n_1, n_2, t) = \frac{1}{\Omega} = 1$, irrespective of the shift, as expected. 

\subsection{Honeycomb Lattice}
Since the honeycomb lattice is created by considering the hexagonal lattice with internal states, the set of images used
to construct the finite hexagonal propagator in Sec. \ref{sec: Periodic} can also be applied to the honeycomb case.
Using Eq. (\ref{eq: honeycomb_unbounded}) in the $6\times 1$ column vectorial equivalent of Eq. (\ref{eq:
periodic_images}), the periodically bounded honeycomb LRW propagator is given by (Appendix \ref{honeycomb-app})
\begin{widetext}
    \begin{equation}\label{eq: honeycomb_periodic}
        \begin{aligned}
            \bm{\mathcal{P}}^{(p)_{[i]}}_{\bm{n}_{0}, m_0}(n_1, n_2, t) = \frac{\bm{\mu}(0,0)^t \cdot \bm{U}_{m_0}}{\Omega} &+ \frac{1}{\Omega}\sum_{r=0}^{R-1}\sum_{s=0}^{3r+2} \Bigg\{e^{\frac{2\pi i(\bm{n} -\bm{n}_0)\cdot\bm{k}^{[i]}(r,s)}{\Omega}}\bm{\mu}\left(\frac{2\pi k_1^{[i]}(r,s)}{\Omega}, \frac{2\pi k_2^{[i]}(r,s)}{\Omega}\right)^t\\ & +
            e^{\frac{-2\pi i(\bm{n} -\bm{n}_0)\cdot\bm{k}^{[i]}(r,s)}{\Omega}}\bm{\mu}\left(\frac{-2\pi k_1^{[i]}(r,s)}{\Omega}, \frac{-2\pi k_2^{[i]}(r,s)}{\Omega}\right)^t \Bigg\}\cdot \bm{U}_{m_0},
        \end{aligned}
    \end{equation}
\end{widetext}
where $k^{[i]}_1(r,s), k^{[i]}_2(r,s)$ are defined in Eq. (\ref{eq: ks}). To lighten the notation, from here onwards we
drop the explicit $(r,s)$ dependence on $k^{[i]}_1(r,s), k^{[i]}_2(r,s)$. 

To obtain $\mathcal{P}^{(p)_{[i]}}_{\bm{n}_{0}, m_0}(n_1, n_2, m, t)$ a scalar dot product is taken, i.e.
$\mathcal{P}^{(p)_{[i]}}_{\bm{n}_{0}, m_0}(n_1, n_2, m, t) =
\bm{U}_{m}^\intercal\cdot\bm{\mathcal{P}}^{(p)_{[i]}}_{\bm{n}_{0}, m_0}(n_1, n_2, t) $. When $R=0$, one is left with six
internal states at the origin and Eq. (\ref{eq: honeycomb_periodic}) reduces to the dynamics dictated by its first term. 

For $0 < q < 1$, as $t \to \infty$, $\bm{\mu}\left(\frac{2\pi k_1^{[i]}}{\Omega}, \frac{2\pi
k_2^{[i]}}{\Omega}\right)^t$, $\bm{\mu}\left(\frac{-2\pi k_1^{[i]}}{\Omega}, \frac{-2\pi k_2^{[i]}}{\Omega}\right)^t \to
\bm{0}$, while $\bm{\mu}\left(0, 0\right)^t \to \frac{1}{6}\mathbb{J}$, where $\mathbb{J}$ is an all-ones matrix (see
Appendix \ref{sec: honey_ss}), leaving the steady state probability as $\mathcal{P}_{\bm{n}_{0}, m_0}^{(p)_{[i]}}(n_1,
n_2, m, t= \infty) = \frac{1}{6\Omega}$. Note that due to the odd coordination number, parity issues appear when $q=1$.
That is, if the walker starts on an even (odd) site number, for large even $t$ the steady state probability on odd
(even) sites $\mathcal{P}_{\bm{n}_{0}, m_0}^{(p)_{[i]}}(n_1, n_2, m, t= \infty) = 0$, while on even (odd) sites,
$\mathcal{P}_{\bm{n}_{0}, m_0}^{(p)_{[i]}}(n_1, n_2, m, t= \infty) = \frac{1}{3\Omega}$. This can again be seen by
studying $\bm{\mu}\left(0, 0\right)$ (Appendix \ref{sec: honey_ss}).

\begin{figure}[h]
	\centering
	\includegraphics[width = 0.34\textwidth]{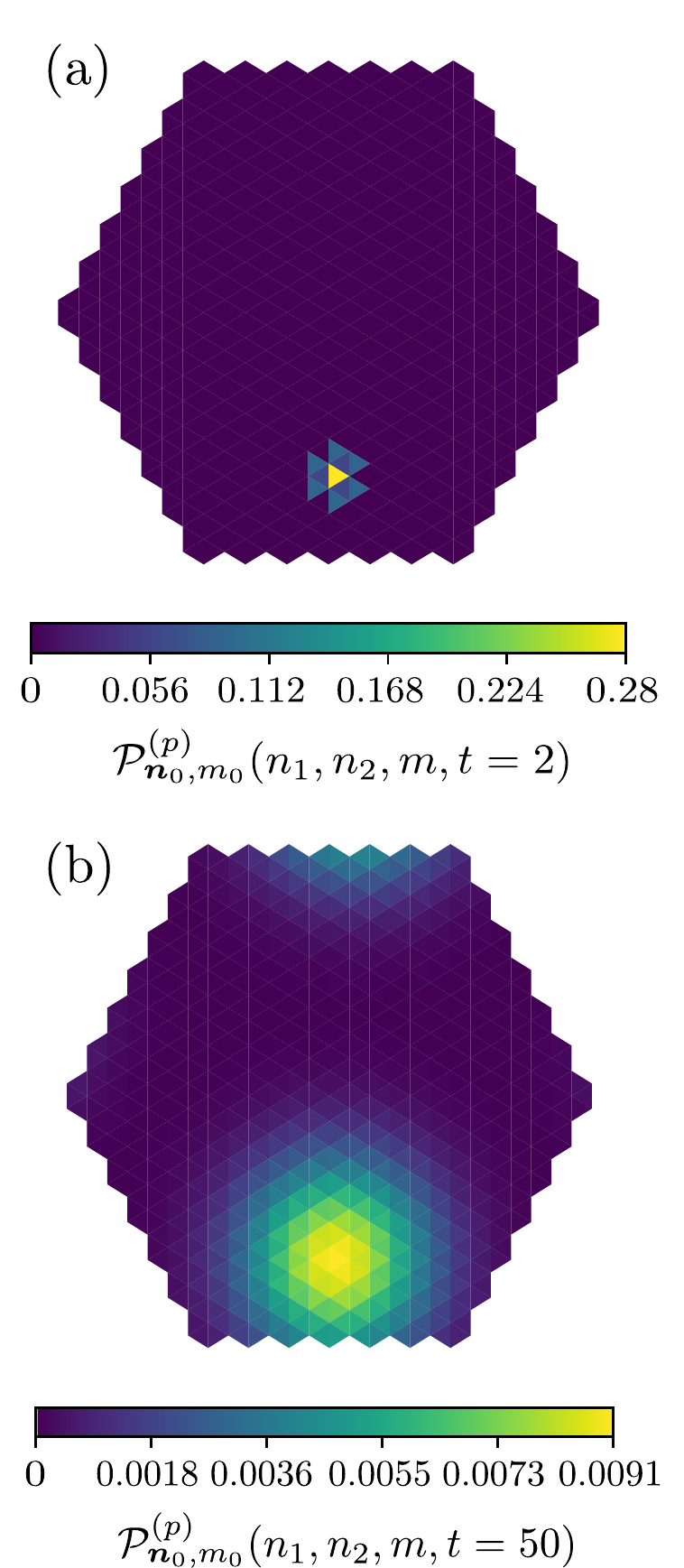}
	\caption{(Colour online). The occupation probability of the right shift bounded periodic honeycomb LRW at $t=2$ in
	(a) and $t=50$ in (b) from Eq. (\ref{eq: honeycomb_periodic}). In both cases, $q=0.9$, $R=6$ and the initial
	condition is at $(n_{0_1}, n_{0_2}, n_{0_3}), m = (2, 2, -4), 4$.}
	\label{fig: honeycomb_temp}
\end{figure}

Note also that the term inside the double summation of Eq. (\ref{eq: honeycomb_periodic}) is the
addition between a matrix and its Hermitian transpose, which ensures the propagator gives real
values. We illustrate this by plotting, from Eq. (\ref{eq: honeycomb_periodic}), the occupation probability after two
separate timesteps in Fig. \ref{fig: honeycomb_temp}. 
\section{Absorbing Boundary Conditions}\label{sec: Absorb} To obtain closed-form solutions with absorbing boundaries we
employ the so-called defect technique \cite{montroll1979enriched, montroll1955effect, luca_transmission}, placing
absorbing defects along boundary sites (or states for the honeycomb lattice). This technique can be applied directly to
the hexagonal lattice, while for the honeycomb lattice, we extend it to random walks with internal states. 

To do this we go beyond the standard way defects are introduced into internal states Master equations. Placing a single
defect across the entire WS cell and modifying the transition probability matrix accordingly in the Master equation
\cite{hughes1983random} is a valid approach but only if the unbounded propagator is chosen as the defect-free one. Here,
we allow any known internal states propagator to be the non-defective solution of the Master equation and place defects
on specific internal states.

For either lattice, we consider the periodic propagator as the defect-free propagator. Since we take the defective
points as fully absorbing, the walker gets taken out of the system upon reaching any boundary point. The absence of any
dynamics in that situation makes the choice of left or right periodic propagator irrelevant. As such, we drop the $\rho,
\lambda$ superscript. 

To obtain temporal dynamics from generating functions, here and elsewhere below, we exploit the convenience of the
numerical inverse $z$-transform \cite{z-inverse}. 

\subsection{Hexagonal Lattice}\label{sec: hex_absorb} We denote the set of boundary points, i.e. the lattice sites with
one or more coordinates equal to $\pm R$, $B^{(a)} = \{\bm{b}_1, \bm{b}_2, ..., \bm{b}_N\}$  where $N = 6R$. One finds
the generating function of the absorbing propagator as \cite{luca_transmission} 
\begin{equation}
\begin{aligned}
    \widetilde{P}_{\bm{n}_{0}}^{(a)}(\bm{n}, z) &= \widetilde{P}_{\bm{n}_{0}}^{(p)}(\bm{n}, z) - \sum_{j = 1}^{N}\widetilde{P}_{\bm{b}_j}^{(p)}(\bm{n}, z)\frac{\det({\mathbb{G}^{(j)}( \bm{n}_{0}, z)})}{\det{(\mathbb{G}(z)})},
\end{aligned}
\label{hex_absorbing_propagator}
\end{equation}
\sloppy where $\mathbb{G}(z)_{i, k} =\widetilde{P}_{\bm{b}_k}^{(p)}(\bm{b}_i, z) $, a $6R\times 6R$ matrix, is built by
considering the defect-free dynamics from one defect to every other defect and $\mathbb{G}^{(j)}(\bm{n}_{0}, z)$ the
same as $\mathbb{G}$ but with the $j^{\mathrm{th}}$ column replaced with the transpose of the vector $\left
[\widetilde{P}_{\bm{n}_{0}}^{(p)}(\bm{b}_1, z),  \widetilde{P}_{\bm{n}_{0}}^{(p)}(\bm{b}_2, z), \dots,
\widetilde{P}_{\bm{n}_{0}}^{(p)}(\bm{b}_M, z)\right]$.

\subsection{Honeycomb Lattice}\label{sec: honey_absorb} We define defective states along the boundary of the honeycomb
lattice $\mathcal{B}^{(a)} =\{ (\bm{b}_1, m_{\bm{b}_1}), (\bm{b}_2, m_{\bm{b}_2}), ...,  (\bm{b}_{\mathcal{N}},
m_{\bm{b}_\mathcal{N}})\}$. The sites in which these defects are placed are equivalent to the hexagonal case. However,
with the inclusion of internal states, the number of defective states is $\mathcal{N} = 6(2R + 1)$, that is two on each
standard boundary site, and three on each corner.

The absorbing propagator for the honeycomb lattice is given as (Appendix \ref{sec: defect_internal_states})
\begin{equation}
\begin{aligned}
    \widetilde{\mathcal{P}}_{(\bm{n}_{0}, m)}^{(a)}&(\bm{n}, m, z) = \widetilde{\mathcal{P}}_{\bm{n}_{0}, m_0}^{(p)}(\bm{n}, m, z) - \\&  \sum_{j = 1}^{\mathcal{N}}\widetilde{\mathcal{P}}_{ (\bm{b}_j, m_j)}^{(p)}(\bm{n}, m, z)\frac{\det({\mathbb{H}^{(j)}( \bm{n}_{0}, m_0, z)})}{\det{(\mathbb{H}(z)})},
\end{aligned}
\label{honey_absorbing_propagator}
\end{equation}
 where $\mathbb{H}(z)_{i,k} = \widetilde{\mathcal{P}}_{(\bm{b}_k, m_{\bm{b}_k})}^{(p)}(\bm{b}_i, m_{\bm{b}_i}, z)$ and
$\mathbb{H}^{(j)}(\bm{n}_{0}, m_0, z)$ is the same as $\mathbb{H}(z)$, but with the $j^{\mathrm{th}}$ column replaced
with the transpose of the vector $\big [\widetilde{\mathcal{P}}_{\bm{n}_{0}, m_0}^{(p)}(\bm{b}_1, m_1, z), \dots,
\widetilde{\mathcal{P}}_{\bm{n}_{0}, m_0}^{(p)}(\bm{b}_M, m_M, z)\big]$.
\section{Reflecting Boundary Conditions}\label{sec: Reflect} 
\begin{figure}[tbhp]
    \centering
    \includegraphics[width = 0.34\textwidth]{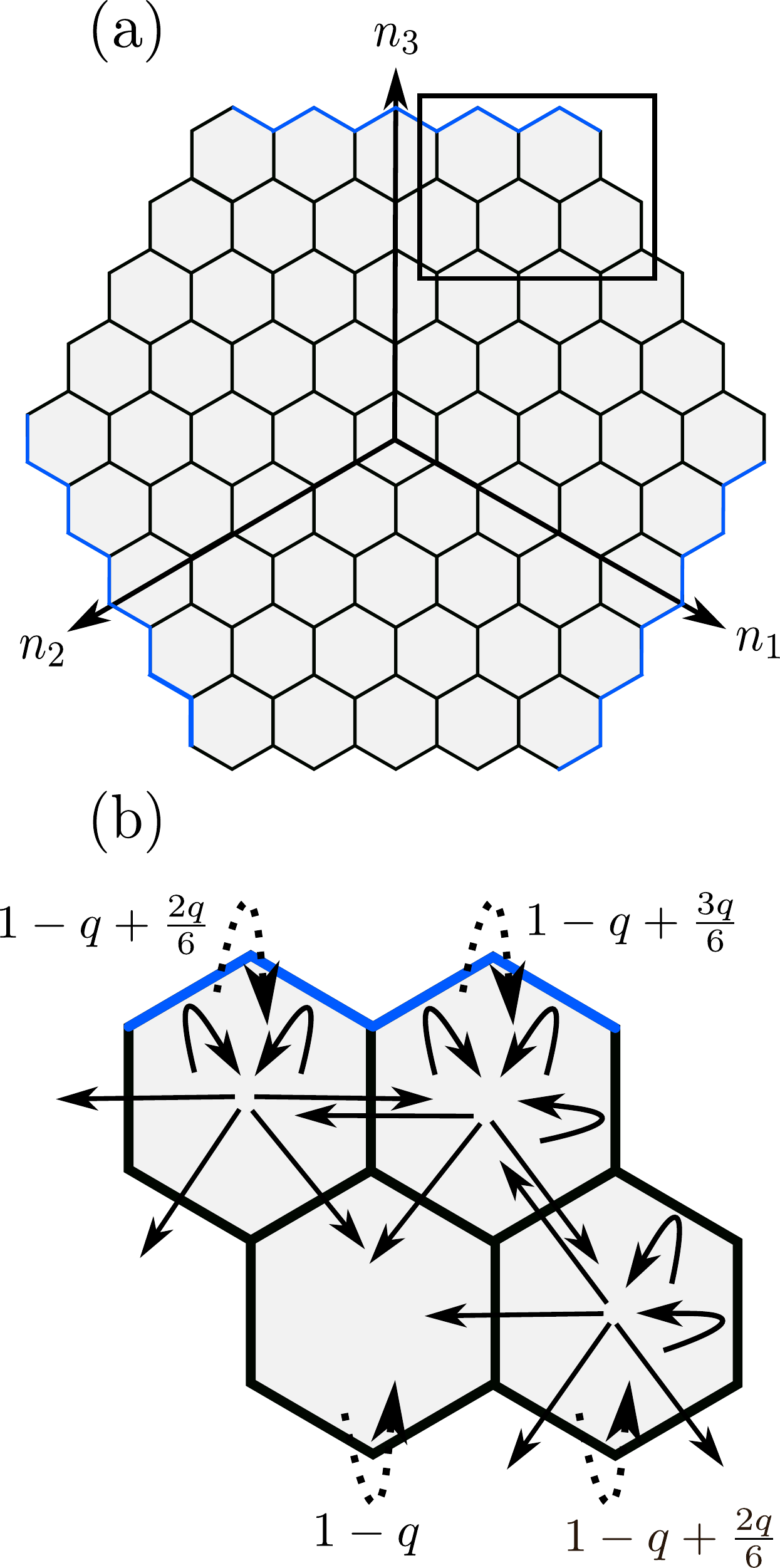}
    \caption{(Colour online). A schematic representation of the reflective hexagonal domain. Panel (a) depicts the
    required barriers to turn a right shift periodic domain into a fully reflective domain. Panel (b) shows the four
    full WS cells inside the rectangle in panel (a). In panel (b) we show how the walker tries to escape and is
    reflected onto the corresponding WS cell. The dotted arrows represent the modified probability of remaining on that
    lattice site and it depends on $q$, as indicated. Note that the probability of remaining on a corner site is
    $\frac{q}{6}$ greater than other boundary sites.}
    \label{fig: reflective_pairs}
\end{figure}

We now place defects between boundary sites (or states).
Taking the periodic propagator as the defect-free solution, we reduce the number of reflective barriers required to make
a fully reflective domain compared to, say, the unbounded propagator. The general formalism derived in
\cite{sree_reflect} can be used in the hexagonal lattice with careful consideration of the placement of reflective
barriers (see Fig. \ref{fig: reflective_pairs}(a)) and we also make it applicable to random walks with internal states
for the honeycomb lattice (see Appendix \ref{app: defect_reflect_internal_states} for derivation). 

Defects between boundary sites (states) are placed by modifying the outgoing connections from boundary site (state)
$\bm{u}$ to boundary site (state) $\bm{v}$ via the parameter $\eta_{\bm{u}, \bm{v}}$, where $0 < \eta_{\bm{u}, \bm{v}}
\leq \frac{q}{Z}$. While it is possible for $\eta_{\bm{u}, \bm{v}} \neq \eta_{\bm{v}, \bm{u}} $ (representing one way or
partial reflection) \cite{sree_reflect}, we take them as equivalent with perfect, bi-directional reflection, i.e.
$\eta_{\bm{u}, \bm{v}} = \eta_{\bm{v}, \bm{u}} = \frac{q}{Z}$ for all boundary interaction in either lattice.

Despite the `zig-zag' boundaries, the movement directions can be thought of as if the walker tries to jump over one of
the boundaries, it gets pushed back in (see Fig. \ref{fig: reflective_pairs}(b) for the hexagonal case) meaning that
reflective dynamics are modelled as if the walker attempts to escape, it remains at the site (or state) it came from. To
find connected boundary sites, one imagines a walker one jump outside the bounded domain and simply subtracts the
coordinate of the centroid of the nearest image to this point. While the following equations do depend on whether the left or
right shift periodic propagator is taken, it is simply a matter of considering the appropriate set of defective sites
(states) for the propagator chosen. Therefore, for ease of notation we drop the $\rho$, $\lambda $ superscripts.

\subsection{Hexagonal Lattice}\label{hex_reflect} We consider the set of defective paired sites $B^{(r)} =
\{\{\bm{b}_1,\bm{b}'_1\},\{\bm{b}_2,\bm{b}'_2\}, ..., \{\bm{b}_{N_1},\bm{b}'_{N_1}\}\}$ where $N_1 = 6R + 3$. Following
\cite{sree_reflect}, and taking $\eta_{\bm{b}_i, \bm{b}'_i} = \eta_{\bm{b}'_i, \bm{b}_i} = \frac{q}{6}$ for all $i$, the
generating function of the propagator is given by
\begin{equation}
\begin{aligned}
    \widetilde{P}_{\bm{n}_{0}}^{(r)}(\bm{n}, z) &= \widetilde{P}_{\bm{n}_{0}}^{(p)}(\bm{n}, z) - 1 + \frac{\det({\mathbb{K}(\bm{n}, \bm{n}_{0}, z)})}{\det{(\mathbb{K}(z)})},
    \label{eq: hex_reflect}
\end{aligned}
\end{equation}
where  $\mathbb{K}(z)$ and $\mathbb{K}(\bm{n}, \bm{n}_{0}, z)$ are $(6R + 3) \times (6R + 3)$ matrices with elements 
\begin{equation}
\begin{aligned}
\mathbb{K}(z)_{i,k} &=\frac{q}{6}\left[\widetilde{P}_{\langle \bm{b}^{\phantom{\prime}}_k - \bm{b}^{\prime}_k\rangle}^{(p)}(\bm{b}_i, z) - \widetilde{P}_{\langle \bm{b}^{\phantom{\prime}}_k - \bm{b}^{\prime}_k \rangle}^{(p)}(\bm{b}^{\prime}_i, z)\right] - \frac{\delta_{i k}}{z},
\end{aligned}
\end{equation}
\begin{equation}
    \begin{aligned}
    \mathbb{K}(\bm{n}, \bm{n}_{0}, z)_{i,k} &= \mathbb{K}(z)_{i,k} - \frac{q}{6}\bigg(\widetilde{P}_{\langle \bm{b}^{\phantom{\prime}}_k - \bm{b}^{\prime}_k\rangle}^{(p)}(\bm{n}, z) \\& \times \left[\widetilde{P}_{\bm{n}_{0}}^{(p)}(\bm{b}^{\phantom{\prime}}_i, z) - \widetilde{P}_{\bm{n}_{0}}^{(p)}(\bm{b}^{\prime}_i, z,)\right]\bigg),
    \end{aligned}
    \end{equation}
respectively, with the notation $f_{\langle \bm{b} - \bm{b}^{\prime}\rangle}(\cdot) = f_{\bm{b}}(\cdot) -
f_{\bm{b}^{\prime}}(\cdot) $.

\subsection{Honeycomb Lattice}\label{sec: honey_reflect} \sloppy The set of pairs of defective states is
$\mathcal{B}^{(r)} = \{\{(\bm{b}^{\phantom{\prime}}_1, m_{\bm{b}^{\phantom{\prime}}_1}), (\bm{b}^{\prime}_1,
m_{\bm{b}^{\prime}_1})\},...,\{(\bm{b}^{\phantom{\prime}}_{\mathcal{N}_1},
m_{\bm{b}^{\phantom{\prime}}_{\mathcal{N}_1}}), (\bm{b}^{\prime}_{\mathcal{N}_1},
m_{\bm{b}^{\prime}_{\mathcal{N}_1}})\}\}$ where the sites $(\bm{b}^{\phantom{\prime}}_1, \bm{b}^{\phantom{\prime}}_2
..., \bm{b}^{\phantom{\prime}}_{\mathcal{N}_1})$ correspond to the defective pairs required for corresponding shift in
the hexagonal case making $\mathcal{N}_1 = N_1$. We adjust the outgoing connections by setting
$\eta_{m_{\bm{b}^{\phantom{\prime}}_i}, m_{\bm{b}^{\prime}_i}} = \eta_{m_{\bm{b}^{\prime}_i}, m_{\bm{b}_i}} =
\frac{q}{3}$. 

The generating function of the propagator is given as (see Appendix \ref{app: defect_reflect_internal_states}) 
\begin{equation}
\label{eq: reflect_honey_prop}
\begin{aligned}
    \widetilde{\mathcal{P}}_{\bm{n}_{0}, m_0}^{(r)}(\bm{n}, m, z) &= \widetilde{\mathcal{P}}_{\bm{n}_{0}, m_0}^{(p)}(\bm{n}, m, z) - 1  \\ &+\frac{\det({\mathbb{L}(\bm{n}, m, \bm{n}_{0}, m_0, z)})}{\det{(\mathbb{L}(z)})},
\end{aligned}
\end{equation}
where  $\mathbb{L}(z)$ and $\mathbb{L}(\bm{n}, m, \bm{n}_{0}, m_0, z)$ are $(6R + 3) \times (6R + 3)$ matrices with
elements 
\begin{equation}
\begin{aligned}
\mathbb{L}(z)_{i,k} &=\frac{q}{3}\bigg[\widetilde{\mathcal{P}}_{\langle \bm{b}_k,m_{\bm{b}_k}   - \bm{b}'_k,m_{\bm{b}'_k}\rangle}^{(p)}(\bm{b}_i, m_{\bm{b}_i}, z) - \\& \widetilde{\mathcal{P}}_{\langle \bm{b}_k,m_{\bm{b}_k}  - \bm{b}'_k, m_{\bm{b}'_k} \rangle}^{(p)}(\bm{b}'_i, m_{\bm{b}'_i}, z)\bigg] -  \frac{\delta_{i k}}{z},
\end{aligned}
\end{equation}
and
\begin{equation}
\begin{aligned}
\mathbb{L}&(\bm{n}, m, \bm{n}_{0}, m_0, z)_{i,k} =  -  \frac{q}{3}\widetilde{\mathcal{P}}_{\langle  \bm{b}_k,m_{\bm{b}_k}  - \bm{b}'_k,m_{\bm{b}'_k}\rangle}^{(p)}(\bm{n}, m, z) \\ &\times \left[\widetilde{\mathcal{P}}_{\bm{n}_{0}, m_0}^{(p)}(\bm{b}^{\phantom{'}}_i, m_{\bm{b}^{\phantom{'}}_i}, z) - \widetilde{\mathcal{P}}_{\bm{n}_{0}, m_0}^{(p)}(\bm{b}^{'}_i, m_{\bm{b}'_i}, z)\right] + \mathbb{L}(z)_{i,k},
\end{aligned}
\end{equation}
respectively.

\section{First-Passage Probability in Periodic Domains}\label{sec: First_Passage} Two important quantities in the
dynamics of stochastic systems are the first-passage, or first hitting, probability $F_{\bm{n}_{0}}(\bm{n}, t)$, and the
return probability $R_{\bm{n}}(t)$. $F_{\bm{n}_{0}}(\bm{n}, t)$ represents the time dependence of the probability to
reach a target $\bm{n}$ from the initial condition $\bm{n}_{0}$, while $R_{\bm{n}}(t)$ represents the first time the
walker returns to $\bm{n} = \bm{n}_{0}$. The generating function of these quantities are obtained via the well-known
renewal equation \cite{rws_on_latticesII}, which is valid in arbitrary dimensions. The generalisation to random walks
with internal states is straightforward leading to, in the $z$-domain, $\widetilde{\mathcal{F}}_{\bm{n}_{0},
m_0}(\bm{n}, m, z) = \widetilde{\mathcal{P}}_{\bm{n}_{0}, m_0}(\bm{n}, m, z)/ \widetilde{\mathcal{P}}_{\bm{n},
m}(\bm{n}, m, z)$ and $\widetilde{\mathcal{R}}_{\bm{n}, m}(z) = 1 - 1/\widetilde{\mathcal{P}}_{\bm{n}, m}(\bm{n}, m, z)$
\cite{rws_on_latticesIII}. In this section, we study the differences between the first-passage temporal dependence of
the left and right shift in periodically bounded domains in the presence of a single target.
\begin{figure*}[]
    \centering 
    \includegraphics[width = 0.8\textwidth]{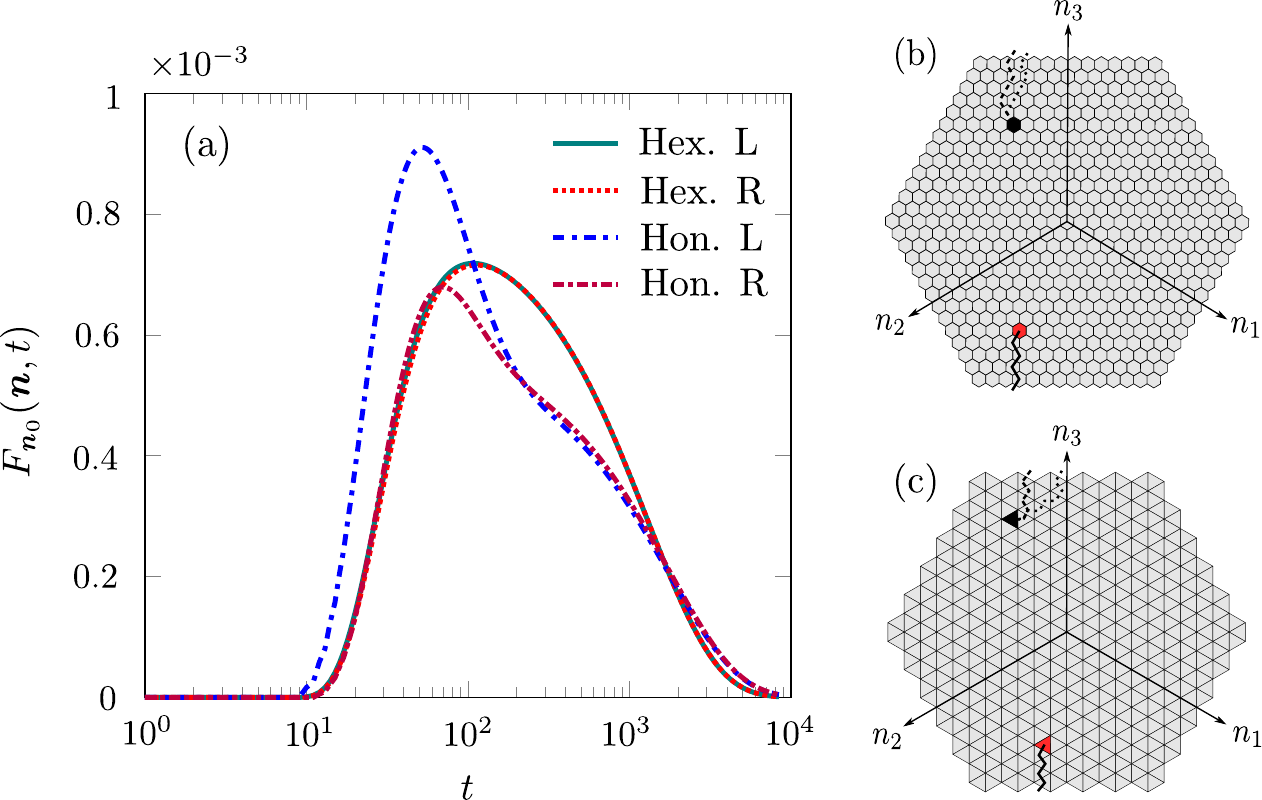}
    \caption{(Colour online). Time dependent first passage probability in periodic hexagonal and honeycomb domains. In
    panel (a) we present the temporal probability for both shifts in both lattices, for the initial condition and
    placement of the targets and initial conditions are presented in panels (b) and (c). The hexagonal lattice, panel
    (b), is $R=13$ (547 lattice sites) and the honeycomb lattice, panel (c), is $R=5$ (546 lattice states). In both
    cases, the initial condition is placed near the bottom left corner, that is $\bm{n}_{0}= (1, 8, -9)$ in the
    hexagonal lattice, and $\bm{n}_{0}, m_0 = (1, 3, -4), 3$ in the honeycomb lattice, while the targets are placed
    across the periodic boundary near the top left corner, at $\bm{n}= (-8, 0, 8)$ in the hexagonal lattice and $\bm{n},
    m = (-4, 0, 4), 3$ in the honeycomb lattice. This ensures that in the hexagonal shift, there is a ballistic
    trajectory of 10 steps for both shifts, while in the honeycomb the left shift has a ballistic trajectory of 10 steps
    and the right shift has a ballistic trajectory of 12 steps. We show one of these trajectories using a solid line
    before the walker crosses the boundary and a dashed line for the left shift and dotted line for the right shift
    after the walker crosses the boundary. For all curves we take $q = 0.85$.}
    \label{fig: FP}
\end{figure*}
It is well known that the direct trajectories, those that travel in a more direct path from the initial condition to the
target, influence the location and the mode of the first-passage probability \cite{godec2016universal}. As the chosen
shift impacts the dynamics at the boundary, if the initial condition and the target are placed across the boundary from
one another, the direct trajectories differ between the two shifts. This difference is greater the smaller the number of
ways to reach the target (or variance in the direct trajectories), which occurs when the locations of the initial
condition and the target lie only a few jumps across a boundary from one another. In this setting, by considering all
the ballistic trajectories, one may expect disparities between the modes of the two shifts' probability as even if there
is an equal number of jumps to reach the target in both cases, one shift may provide more ballistic options than the
other.

In Fig. \ref{fig: FP}, we show one such case by placing the walkers with identical initial conditions, near the bottom
left of the domains, and placing the targets across the boundary near the top left corner (see Fig. \ref{fig: FP} panels
(b), (c)). In this setting, the lower coordination number for the honeycomb lattice ensures a lower variance in the
direct trajectories. In turn, this causes the trajectories that differ between the shifts to have a greater effect on
the modes of the distribution. While the difference in first-hitting dynamics is evident for the honeycomb case, to
observe similar disparities for the hexagonal lattice one needs to move the initial condition and target closer to the
boundary. 

Despite the mode dynamics being considerably different in the honeycomb lattice, the tails of the distribution are very
similar as the tail is heavily dependent on the indirect trajectories \cite{godec2016universal}, i.e. the paths where
the walker meanders around the domain and does not hit the target for extended periods of time. In the honeycomb case we
see clearly when the indirect trajectories become dominant as it corresponds to the kinks in the temporal dependence at
around $t\approx110$.

Analytic knowledge of the propagators also allows us to readily calculate the first passage to a set of $\mathcal{M}$
targets $\{S\} = \{\bm{s}_1,...,\bm{s}_{\mathcal{M}}\}$. This is done by considering the splitting probabilities, that
is the probability of reaching one target $\bm{s}_j$ in $\{S\}$ before reaching any other, which is given by
\cite{luca_transmission}
\begin{equation}
\label{eq: splitting}
\widetilde{T}_{\bm{n}_{0}\to (\bm{s}_j|\{S\}- \bm{s}_j) }(z) = \frac{\det(\mathbb{F}^{(j)}(\bm{n}_0, z))}{\det(\mathbb{F}(z))},
\end{equation}
where $\mathbb{F}(z)_{k,k} = 1$, $\mathbb{F}(z)_{i, k} =
\widetilde{F}_{\bm{s}_k}(\bm{s}_i, z)$ and $\mathbb{F}^{(j)}(\bm{n}_0, z)$ is the same as $\mathbb{F}(z)$ but with the
$j^{\text{th}}$ column replaced with $\left [\widetilde{F}_{\bm{n}_{0}}(\bm{s}_1, z),
\widetilde{F}_{\bm{n}_{0}}(\bm{s}_2, z), \dots, \widetilde{F}_{\bm{n}_{0}}(\bm{s}_{\mathcal{M}},
z)\right]^{\intercal}$.  Since the splitting probabilities represent mutually exclusive trajectories, to obtain the
generating function for the first-passage to any target, one simply sums them, i.e. $ \widetilde{T}_{\bm{n}_{0}\to
\{S\}} = \sum_{j = 1}^{\mathcal{M}} \widetilde{T}_{\bm{n}_{0}\to (\bm{s}_j|\{S\}- \bm{s}_j) }(z)$.
For all known internal states propagators with localised initial conditions, the analogous quantities to those in Eq.
(\ref{eq: splitting}) are easily found, ensuring a trivial extension to the honeycomb lattice.

\section{Mean First Passage Time} \label{sec: MFPT} Analytic knowledge of the generating functions of the first-passage
and return probabilities allow us to obtain closed-form representation of their first moments
($\mathdutchcal{F}_{\bm{n}_0 \to \bm{n}}$ and $\mathdutchcal{R}_{\bm{n}_0}$) in the hexagonal and honeycomb lattices
found by evaluating the first derivative, with respect to $z$, of the respective probability generating function at
$z=1$ \cite{rws_on_latticesII}. The MFPT to multiple targets is also accessible via
\begin{equation}
	\label{eq: mean_transmission}
	\mathdutchcal{T}_{\bm{n}_{0} \to \{S\}} = \frac{\det(\mathbb{T}_0)}{\det(\mathbb{T}_1) - \det(\mathbb{T})},
\end{equation}
a general result derived more recently in \cite{luca_transmission}, where $\mathbb{T}_{ii} =
0$,  $\mathbb{T}_{ij} =\mathdutchcal{F}_{\bm{s}_j \to \bm{s}_i}$, $\mathbb{T}_{0_{ij}} = \mathbb{T}_{ij} -
\mathdutchcal{F}_{\bm{n}_{0}\to\bm{s}_i}$ and $\mathbb{T}_{1_{ij}} = \mathbb{T}_{ij} - 1$.
\subsection{Hexagonal Lattice}
For the hexagonal lattice, the MFPT is given by
\begin{figure*}
    \centering 
    \includegraphics[width = 0.8\textwidth]{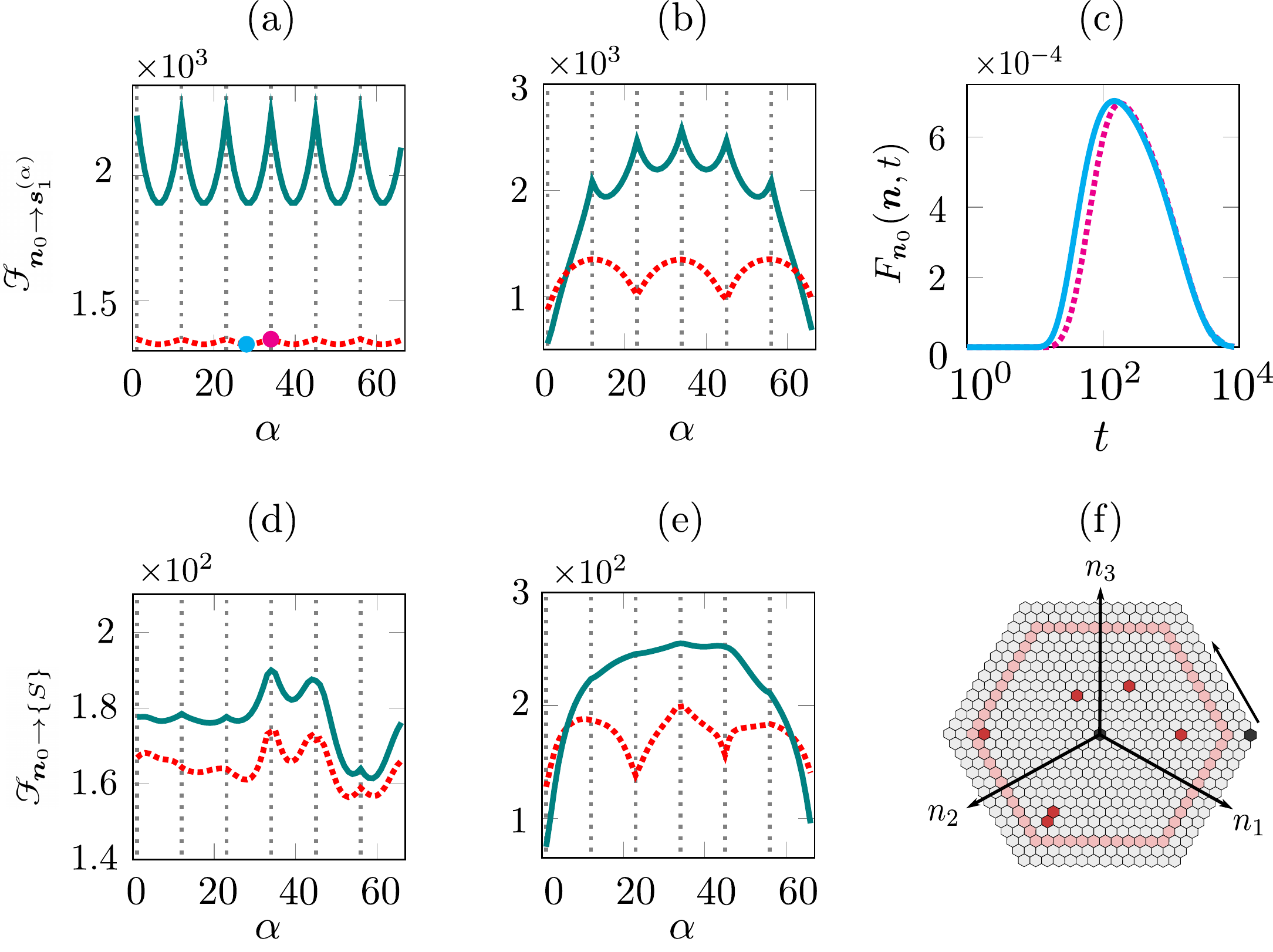}
    \caption{(Colour online). Panels (a), (b), (d) and (e) represent the MFPT in periodic (dashed lines) and reflective
    (solid lines) with $R = 13$ hexagonal lattices. In (a) and (d), $\bm{n}_{0} = (0, 0, 0)$, while in (b) and (e)
    $\bm{n}_{0} = (13, -13, 0)$. The upper two panels correspond to a single target system $\bm{s}^{(\alpha)}_1$, i.e.
    we use Eqs. (\ref{eq: periodic_MFPT}) and (\ref{eq: reflect_MFPT}), while for the bottom two panels, we have a
    system with seven targets, i.e. we use Eq. (\ref{eq: mean_transmission}), retaining $\bm{s}^{(\alpha)}_1$ and
    introducing six other targets at $\bm{s}_{\{2,..7\}} = \{(-10, 10, 0), (0, 9, -9), (0, 8, -8), (7, -7, 0), (0, -5,
    5), (-4, 0, 4)\}$ giving the set $\{S\}$ targets. In both cases, we move $\bm{s}^{(\alpha)}_1$ anti-clockwise around
    the $R = 11$ ring of coordinates where $\bm{s}^{(1)}_1 = (11, -11, 0)$, the right most point of the ring,
    $\bm{s}^{(2)}_1 = (10, -11, 1)$ and so on until $\bm{s}^{(65)}_1 = (11, -10, -1)$. Thus, in total $s^{(\alpha)}_1$
    moved through 65 locations. We show, in panel (f), the placement of these six additional targets (dark red), the
    ring (light red) with the change in the placement of the target on the ring shown by the arrow and the two initial
    conditions (black). The vertical grey lines in panels (a), (b), (d) and (e) denote the corner points of this ring.
    In panel (c), we show the temporal first-passage dependence in a right shift periodic domain from $\bm{n}_{0} = (0,
    0, 0)$ to $\bm{s}^{(28)}_1 = (-11, 11, 0)$ (left dot in panel (a)) and $\bm{s}^{(34)}_1 = (-11, 5, 6)$ (right dot in
    panel (a)), shown in, respectively, dashed and solid lines. For all plots we take $q=6/7$.}
    \label{fig: MFPT-comparison}
\end{figure*}
\begin{widetext}
    \begin{equation}
 \label{eq: periodic_MFPT}
	\begin{aligned}
	 \mathdutchcal{F}_{\bm{n}_0\to \bm{n}}^{(p)_{[i]}}  =  \frac{2}{q}\sum_{r=0}^{R-1}&\sum_{s=0}^{3r+2}\left\{ \cos \left( \frac{2\pi k^{[i]}_1(n_1-n_{0_1})+ 2\pi k^{[i]}_2(n_2-n_{0_2})}{\Omega} \right) - 1 \right\}   \\
	 &\times \left\{  \frac{1}{3}\left [\cos \left(\frac{2\pi(k_1^{[i]}-k_2^{[i]})}{\Omega}\right) + \cos\left(\frac{2\pi k_1^{[i]}}{\Omega}\right) + \cos\left(\frac{2\pi k_2^{[i]}}{\Omega} \right) \right] -1\right\}^{-1},
	\end{aligned}
\end{equation} 
\end{widetext}
while for the MRT we confirm Kac's lemma \cite{kac1947notion}, for which, regardless of the shift,
$\mathdutchcal{R}^{(p)}_{\bm{n}} = \Omega = 3R^2 + 3R + 1$, the inverse of the steady state probability.

Knowledge of the generating function of the reflective propagator allows us to study the same statistics in reflective
domains \cite{sree_reflect}. The MFPT is given by
\begin{equation}
	\label{eq: reflect_MFPT}
	\mathdutchcal{F}^{(r)}_{\bm{n}_{0}\to \bm{n}} = \mathdutchcal{F}^{(p)}_{\bm{n}_{0}\to \bm{n}} - 1 + \frac{\det(\mathbb{F}-\mathbb{F}^{(1)})}{\det(\mathbb{F})},
\end{equation}
where 
\begin{equation}
	\label{eq: MFPT_mat1}
	\mathbb{F}_{ij} = \frac{q}{6\Omega}\left[\mathdutchcal{F}^{(p)}_{\langle \bm{b}_j - \bm{b}'_j \rangle \to \bm{b}_i } - \mathdutchcal{F}^{(p)}_{\langle \bm{b}_j - \bm{b}'_j \rangle \to \bm{b}'_i }\right] + \delta_{ij},
\end{equation}
and 
\begin{equation}
\label{eq: MFPT_mat2}
	\mathbb{F}^{(1)}_{ij} = \frac{q\mathdutchcal{F}^{(p)}_{\langle \bm{b}_j - \bm{b}'_j \rangle \to \bm{n}} }{6\Omega}\left[\mathdutchcal{F}^{(p)}_{\langle \bm{n}_{0} - \bm{n} \rangle \to \bm{b}_i } - \mathdutchcal{F}^{(p)}_{\langle \bm{n}_{0} - \bm{n} \rangle \to \bm{b}'_i }\right],
\end{equation}
while the MRT is $\mathdutchcal{R}^{(r)}_{\bm{n}} = \mathdutchcal{R}^{(p)}_{\bm{n}} $, as expected.

In Fig. \ref{fig: MFPT-comparison} we plot the MFPT as a function of the target location for two different initial
conditions $\bm{n}_{0} = (0,0,0)$ (panels (a) and (d)) and $\bm{n}_{0} = (13, -13, 0)$, the far right corner, (panels
(b) and (e)). The target, $\bm{s}^{(\alpha)}_1$, is placed sequentially in a ring-like manner anti-clockwise around the
$R=11$ circumradius of the hexagon. For the case where $\bm{n}_{0} = (0,0,0)$, although the initial displacement between
the initial condition and any target is constant, rich dynamics appear the target is moved around the ring. 

Owing to the symmetry of the system, in both domains, we see oscillations occurring with a wavelength of 11, the length
of a side of the ring the target is moving around. Peaks are located at the corners of the $R=11$ circumradius, while
the troughs correspond to the centre of the ring. At short times, it is easier for a walker to hit a target in the
centre of the ring than the corner as the number of direct trajectories is greater for the centre target, seen in the
modes of Fig. \ref{fig: MFPT-comparison}(c). Owing to the probability conserving property, the tail of
$F_{\bm{n}_{0}}(\bm{n}, t)$ is then slightly lower, giving a smaller MFPT. Furthermore, the MFPT in the reflective cases
is roughly twice that of the corresponding periodic case. This can be understood by thinking that the periodic boundary
conditions effectively double the trajectories with which the walker can reach a target compared to the reflective case. 

In panel (b) there is a marked difference between the dynamics in the reflective and periodic domains. As $\bm{n}_{0}$
is the far right point of the domain, the displacement between the initial condition and a target at, for example,
$\bm{s}^{(23)}_1 = (-11, 0, 11)$ is much greater in the reflective domain than in the periodic domain.  In the
reflective case, we see a linear increase (decrease) as we move the target further away from (closer to) the initial
condition. The linear increase is seen until the target moves around the first corner. We then see small oscillations
where, again, targets at the centre of the ring produce a local minimum and the peaks are located at the corners. The
highest peak corresponds to the target at $\bm{s}^{(34)}_1 = (-11, 11, 0)$, the target furthest from the initial
condition.

We now introduce six other static targets $\{\bm{s}_2, ... , \bm{s}_6\}$ placed at other locations within the $R=11$
ring (given explicitly in the caption of Fig. \ref{fig: MFPT-comparison}) and move $\bm{s}^{(\alpha)}_1$ sequently as
before. For the $\bm{n}_{0} = (0, 0, 0)$ case, the introduction of more targets minimises the differences between the
two boundaries compared to the one target set-up. This is likely due to the targets being placed across the whole domain
meaning for many realisations, the walker will hit a target before any boundary interaction. In contrast, for
$\bm{n}_{0} = (13, -13, 0)$, we see similar behaviour to the one target case as in the reflective case, the initial
condition renders targets on the opposite side of the domain nearly redundant as many other targets lie between them and
the initial condition.
\begin{figure*}
    \includegraphics[width = \textwidth]{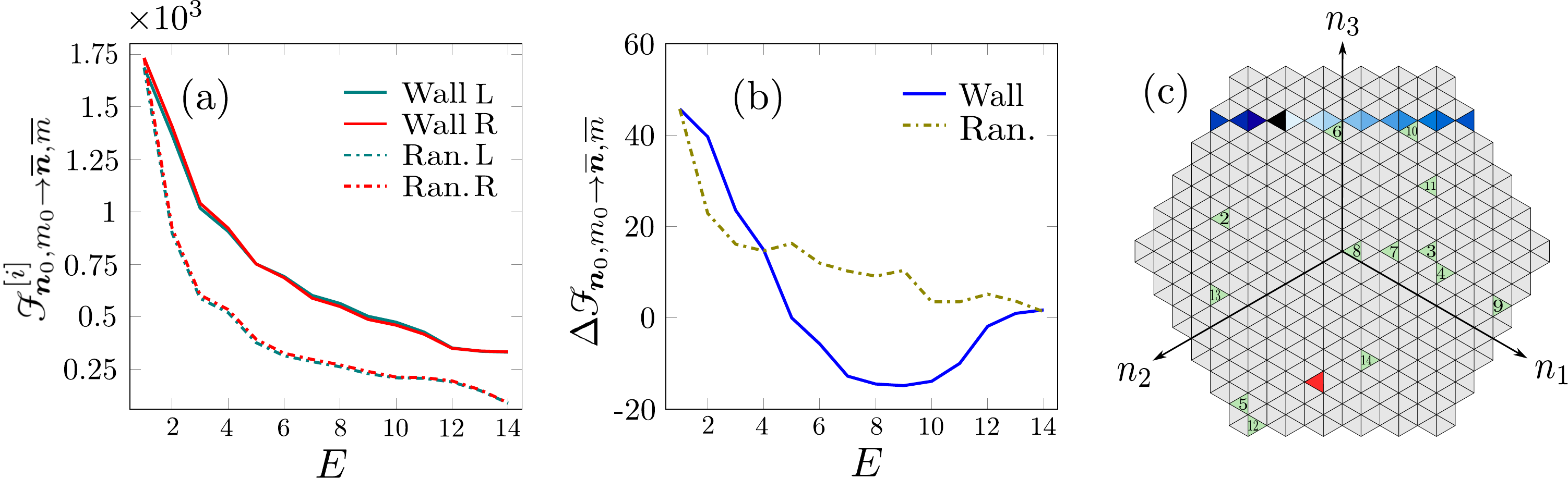}
    \caption{(Colour Online). The MFPT in the left and right shift periodic honeycomb domains as a function of the
            number of targets, $E$, for a randomly placed set of targets (dotted lines) and a wall of targets (solid
            line). Panel (a) represents the MFPT to $E$ targets, $ \mathdutchcal{F}^{[i]}_{\bm{n}_0, m_0 \to \overline{\bm{n}},
            \overline{m}}$, with the bar notation denoting the option of multiple targets. Panel (b) shows the difference, $\Delta\mathdutchcal{F}_{\bm{n}_0, m_0 \to
            \overline{\bm{n}}, \overline{m}} = \mathdutchcal{F}^{[\rho]}_{\bm{n}_0, m_0 \to \overline{\bm{n}},
            \overline{m}}  - \mathdutchcal{F}^{[\lambda]}_{\bm{n}_0, m_0 \to \overline{\bm{n}}, \overline{m}} $, between
            the two shifts. We show the order and placement of the the targets in panel (c). The first target at
            $\bm{n}, m = (-4, 0, 4), 3$, is shared by both set-ups and we denote its place in black. The other targets
            are added sequentially, which we show via the blue colour gradient from light to dark in the wall case and
            numerically for the random case, with targets shown in green. The initial condition at $\bm{n}_0, m_0 = (1,
            3, -4), 3$ is shown in red and $q = 0.85$. }
    \label{fig: honey_MFPT}
\end{figure*}
For both initial conditions, we again see a maxima at $\alpha = 34$, the location where $\bm{s}^{(34)}_1$ is located
next to $\bm{s}_2$. As such, it renders one of these targets near redundant as if the walker is to find one of the
targets, it is likely he will find both. In contrast, the shortest MFPT is when $\bm{s}^{(\alpha)}_1$ is dependent on
the initial condition. For the case with the origin at the initial condition, the minima for both reflective and
periodic is located near the centre of the bottom right side of the domain. This placement of $\bm{s}^{(\alpha)}_1$
fills the space and creates the most widely spread arrangement of targets meaning that a walker exploring any section of
the domain is always in close proximity to a target. For $\bm{n}_0 = (13, -13, 0)$, the effect of the boundary
conditions are still visible with multiple targets. In both domains, the minima naturally lie where the moving target is
only a few jumps from the initial condition.

\subsection{Honeycomb Lattice}
For the honeycomb lattice, we find the MFPT as
\begin{widetext}
    \begin{equation}
        \begin{aligned}
        &\mathdutchcal{F}^{(p)_{[i]}}_{\bm{n}_0, m_0 \to \bm{n}, m} = \bm{U}_{m}^{\intercal}\cdot\mathbb{C}\cdot\bm{U}_{m_0} - \bm{U}_{m}^{\intercal}\cdot\mathbb{C}\cdot\bm{U}_{m} + \\ &
        \bm{U}^{\intercal}_{m}\cdot\Bigg[ 6\sum_{r=0}^{R-1}\sum_{s=0}^{3r+2} \Bigg\{e^{\frac{2\pi i(\bm{n} -\bm{n}_0)\cdot\bm{k}^{[i]}}{\Omega}}\left[\bm{\mu}\left(\frac{2\pi k_1^{[i]}}{\Omega}, \frac{2\pi k_2^{[i]}}{\Omega}\right)-\mathbb{I}\right]^{-1} 
        +  e^{\frac{-2\pi i(\bm{n} -\bm{n}_0)\cdot\bm{k}^{[i]}}{\Omega}}\left[\bm{\mu}\left(\frac{-2\pi k_1^{[i]}}{\Omega}, \frac{-2\pi k_2^{[i]}}{\Omega}\right)- \mathbb{I}\right]^{-1} \Bigg\}\Bigg]\cdot \bm{U}_{m_0}  \\ &-
        \bm{U}^{\intercal}_{m}\cdot\bigg[6\sum_{r=0}^{R-1}\sum_{s=0}^{3r+2} \Bigg\{\left[\bm{\mu}\left(\frac{2\pi k_1^{[i]}}{\Omega}, \frac{2\pi k_2^{[i]}}{\Omega}\right)- \mathbb{I}\right]^{-1} +
       \left[\bm{\mu}\left(\frac{-2\pi k_1^{[i]}}{\Omega}, \frac{-2\pi k_2^{[i]}}{\Omega}\right)- \mathbb{I}\right]^{-1} \Bigg\}\Bigg]\cdot \bm{U}_{m} 
        \end{aligned}
        \label{eq: honey_MFPT}
    \end{equation}
\end{widetext} 
where $\mathbb{C}$ is a $6\times 6$ symmetric circulant matrix with the first row $c = \left[5-\frac{9}{q},
5-\frac{4}{q}, 5-\frac{3}{q}, 5-\frac{4}{q}, 5-\frac{3}{q}, 5-\frac{4}{q} \right]$ such that $\mathbb{C}_{ij} = c_{j-i
\Mod{6}}$. The term $\bm{U}_{m}^{\intercal}\cdot\mathbb{C}\cdot\bm{U}_{m_0} -
\bm{U}_{m}^{\intercal}\cdot\mathbb{C}\cdot\bm{U}_{m}$ is independent of the choice of shift and governs the MFPT in the
degenerate $R=0$ case, which is periodic in $|m-m_0|$ and given explicitly in Eq. (\ref{eq: R0_MFPT1}). It gives the
MFPT for any $m, m_0$ pairs that are either one jump away or those that are two jumps away from one another (see Fig.
\ref{fig: hex_honey_coords}b when R=0 for visual understanding). We again confirm Kac's lemma as the MRT is found to be
$\mathdutchcal{R}^{(p)}_{\bm{n}_0, m_0} = 6\Omega$, the number of states in the lattice. For the reflective case, one
obtains similar expressions to the hexagonal lattice with reflecting boundaries (see Appendix \ref{app: ref_honey_MFPT}).

Using Eq. (\ref{eq: honey_MFPT}) we study the differences between the MFPT of the two shifts as a function of the number
of targets. We sequentially introduce targets in two ways, either building a `wall' of targets along the top of the
domain or placing targets at random locations, with both set-ups shown pictorially in panel (c) of Fig. \ref{fig:
honey_MFPT}. In both the random and the `wall' placement of targets, the initial condition and the location of the first
target correspond to the set-up used to obtain the full FP probability given in Fig. \ref{fig: FP}.

As Fig. \ref{fig: honey_MFPT}, Panel (a) shows, for both target set-ups, adding targets lowers the MFPT, as expected. In
the randomly placed target case, an increase in the number $E$ of targets reduces the differences between the MFPTs, if
the added target is placed in the bulk of the domain, seen in Fig. \ref{fig: honey_MFPT}, Panel (b). This reduction is
due to the increasing likelihood that the walker reaches the target before any boundary interactions occur. If, on the
other hand, the new target is placed on, or very closed to, the boundary ($E = 5$, $E = 9$ and $E = 12$) slight
increases are seen in $\Delta \mathdutchcal{F}_{\bm{n}_0, m_0 \to \bm{n}, m}$ further emphasising the importance of
boundary interactions in the periodic propagators.

In the case of the `wall', we add the targets to the right of the initial target until we reach it again from the left.
In this case we see a dramatic convergence between the two shifts before the mean of the right shift becomes lower. With
a wall of targets placed near the top of the domain, the easiest route for the walker to complete its search is the few
steps across the boundary. As exemplified by the higher first-passage probability mode of the left shift honeycomb walker (Fig. \ref{fig: FP}), when there are few targets, the searcher benefits from the left
shift boundary condition since the distance to the target is smaller. However, as we build the wall to the right, this advantage
lessens until sufficient targets are added such that it is more beneficial to utilise the right shift walker. Finally, when the
wall is complete ($E = 14$) we see negligible differences between the two shifts.
\section{Conclusions} \label{sec: conclusions} 
The implementation of the method of images to obtain periodically bounded LRW in square geometries has been known for
some time \cite{barry_hughes_book,montroll1979enriched}. Somewhat surprisingly, the same could not be said about
hexagonal lattices. Here, we have constructed the image set for a lattice in HCC and derived the exact spatio-temporal dynamics
for a LRW on a periodic hexagonal and honeycomb lattice. 
By generalising the defect technique to hexagonal geometries 
we have found the absorbing and reflecting propagators for both lattices. We 
have then utilised these propagators to obtain expressions such as the return and first-passage
probabilities and their means. 

We note that while we limit ourselves to deploying the defect technique for the dynamics in hexagonally constrained
spatial domains, it is possible to place both absorbing or reflecting sites in such a way as to confine the domain to
other shapes, for example, a triangle. Moreover, the formalism may be applied to other periodic propagators in hexagonal
geometries, for example, a biased LRW \cite{sarvaharman2020closed}, a resetting random walker \cite{das2022discrete} or
when the space is composed of different media \cite{das2023dynamics}. Dynamics on other lattice geometries are also
available through our internal states procedure. For example, by placing three internal states in a triangular
structure, one can create the so-called tri-hexagonal lattice \cite{nagy2019continuous} or with four internal states,
one can achieve a square-octagon tessellation seen in the theorised T-graphene structure \cite{liu2012structural}. Other
potential directions include finding continuum limits of the periodic propagator and placing both absorbing defects
\cite{kenkre2021memory} or reflecting defects \cite{kay2022diffusion} to obtain diffusive dynamics in hexagonally shaped
domains, avoiding the need to solve the diffusion equation numerically in these geometries. We conclude by noting other
potential applications of our work. These include modelling neutron diffusion in a nuclear reactor core
\cite{gonzalez2010time, HEBERT2008363}, the transmission of an infectious pathogen in a population of territorial
animals \cite{robles2018phase, giuggioli2014consequences, sarvaharman2019micro, kenkre2021theory}, diffusion on a SWCN with topological defects
such as dislocations \cite{charlier2002defects}, and amoeboid migration in Petri dishes with hexagonally placed
micropillars \cite{gorelashvili2014amoeboid}. 

\begin{acknowledgments}
    LG acknowledges funding from the Biotechnology and Biological Sciences Research Council (BBSRC) Grant No.
    BB/T012196/1 and the Natural Environment Research Council (NERC) Grant No. NE/W00545X/1, while DM and SS acknowledge
    funding from Engineering and Physical Sciences Research Council (EPSRC) DTP studentships with Reference Nos. 2610858
    and 2123342, respectively.    
\end{acknowledgments}
\onecolumngrid
\appendix

\section{Placement of Images}\label{app: image placement} 
\begin{figure*}[h]
	\centering
	\includegraphics[width = 0.75\textwidth]{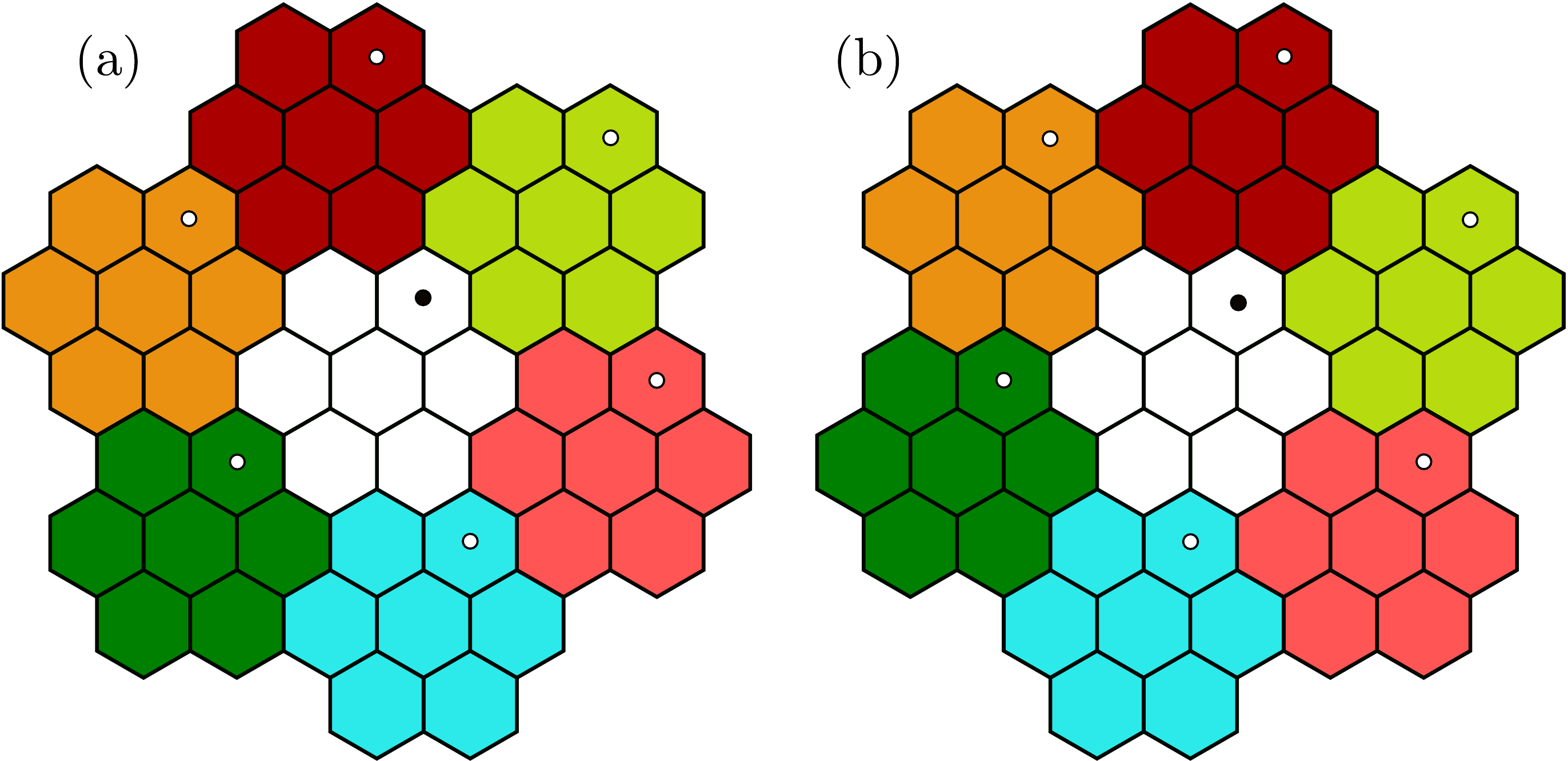}
	\caption{(Colour online). A schematic representation of the nearest-neighbour images for panel (a), the left shift,
	and panel (b), the right shift, in an $R=1$ domain. In this case, we show in the central domain, the LRW starting
	location as a filled circle while the first ring of images has open circles. For ease of visual comparison between
	the left and right shift, each hexagon in the nearest-neighbour image ring is coloured differently.}
	\label{fig: periodic_images}
\end{figure*}
Figure \ref{fig: periodic_images} shows the first ring of the
infinite images for an $R=1$ domain as per discussion in Sec. \ref{sec: Periodic}. We refer to the shift with respect to
the location of the top red image, that is whether it is to the left or the right of the main domain. 

\section{Derivation of the Periodic Hexagonal Propagator}\label{app: periodic} Using the periodic image set, Eqs.
(\ref{eq: right_shift1}) and (\ref{eq: left_shift1}), in the unbounded Green's lattice function, Eq. (\ref{eq:
unbounded_prop}) in the main text, and isolating the image contribution in the exponential, we obtain
\begin{equation}
    \label{eq: bounded_prop2}
    \begin{aligned}
        \widetilde{P}^{[i]}_{\bm{n}_{0}}(n_1, n_2, z) = \frac{1}{(2\pi)^2}\int_{-\pi}^{\pi} \int_{-\pi}^{\pi} \sum_{m_1 = -\infty}^{\infty}\sum_{m_2 = -\infty}^{\infty}  
        \frac{e^{i[(\bm{n} - \bm{n}_0)\cdot\bm{k}]} e^{i[\bm{m} \cdot \mathbb{D}_{[i]}\cdot \bm{k}]}} {1-z\mu(k_1, k_2)}dk_1 k_2,
    \end{aligned}
    \end{equation}
where $\bm{m} = (m_1, m_2)$, and for the right shift
\begin{equation}
	\label{omega_right_definition}
	\mathbb{D}_{[\rho]} = 
		\begin{bmatrix}
			-R & -R-1 \\
			2R+1 & -R
		\end{bmatrix},
\end{equation}
while for the left shift
\begin{equation}
	\label{omega_left_definition}
	\mathbb{D}_{[\lambda]} = 
		\begin{bmatrix}
			2R + 1 & -R-1 \\
			-R & 2R+1
		\end{bmatrix}.	
\end{equation}
Equations (\ref{eq: bounded_prop2}), (\ref{omega_right_definition}), and (\ref{omega_left_definition}) allow us to
connect with literature on Fourier analysis in hexagonal domains \cite{li2008discrete, cube_hex_coords} and utilise the
distributional form of the Poisson summation formula associated with the hexagonal lattice
\begin{equation}
	\label{eq: poisson_summation}
    \begin{aligned}
        \sum_{m_1 = -\infty}^{\infty}\sum_{m_2 = -\infty}^{\infty}e^{i[\bm{m} \cdot \mathbb{D}_{[i]}\cdot \bm{k}]} = \frac{(2\pi)^2}{\Omega}\sum_{m_1 = -\infty}^{\infty}\sum_{m_2 = -\infty}^{\infty}\delta(\bm{k} - 2\pi\mathbb{D}_{[i]}^{-1}\cdot\bm{m}^{\intercal}),
    \end{aligned}
	\end{equation}
where $\Omega = 3R^2 + 3R +1$, the number of sites in the hexagonal lattice. The equivalent result for the square
lattice can be found in \cite{lighthill_1958}. We note here the importance of dropping the $n_3$ dependence from Eq.
(\ref{eq: original-master-eq}). If the full three co-ordinate representation of HCC was chosen, $\mathbb{D}_{[\rho,
\lambda]}$ would be a $3\times3$ matrix, with three linearly dependent rows making
$\mathbb{D}_{[i]}$ singular matrices.

Applying Eq. (\ref{eq: poisson_summation}) on Eq. (\ref{eq: bounded_prop2}) and shifting the integral limits, due to the
periodicity of the integrand, we obtain  
\begin{equation}
    \begin{aligned}
    \widetilde{P}^{[i]}_{\bm{n}_{0}}(n_1, n_2, z) = \int_{ - \varepsilon}^{2\pi - \varepsilon} \int_{ - \varepsilon}^{2\pi - \varepsilon} \sum_{m_1 = -\infty}^{\infty}\sum_{m_2 = -\infty}^{\infty} \frac{e^{i[(\bm{n} - \bm{n}_0)\cdot\bm{k}]}\delta(\bm{k} - 2\pi\mathbb{D}_{[i]}^{-1}\cdot\bm{m}^{\intercal})} {\Omega\left[1-z \mu(k_1, k_2)\right]}dk_1 dk_2,
    \end{aligned}
    \label{eq: bounded_prop3}
\end{equation}
where the parameter $0 < \varepsilon \leq \frac{2\pi}{\Omega}$, avoids having a singularity of the Dirac delta on the
integral bound. The values where the Dirac delta is non-zero are given by 
\begin{equation}
	\label{eq: parameterise_ksR}
    \begin{aligned}
        k^{[\rho]}_1 &= \frac{2\pi}{\Omega}[-Rm_1 + (R+1)m_2], \\ k^{[\rho]}_2 &= \frac{-2\pi}{\Omega}[(2R+1) m_1 + Rm_2],
    \end{aligned}	
\end{equation}
for the right shift and 
\begin{equation}
	\label{eq: parameterise_kL}
    \begin{aligned}
	k^{[\lambda]}_1 &= \frac{2\pi}{\Omega}[(2R+1)m_1 + (R+1)m_2], \\ k^{[\lambda]}_2 &= \frac{2\pi}{\Omega}[R m_1 + (2R+1)m_2],
    \end{aligned}
\end{equation}
for the left shift. 

\begin{figure}
\centering
\subfloat[]{\label{fig: leftks}\includegraphics[width = 0.42\textwidth]{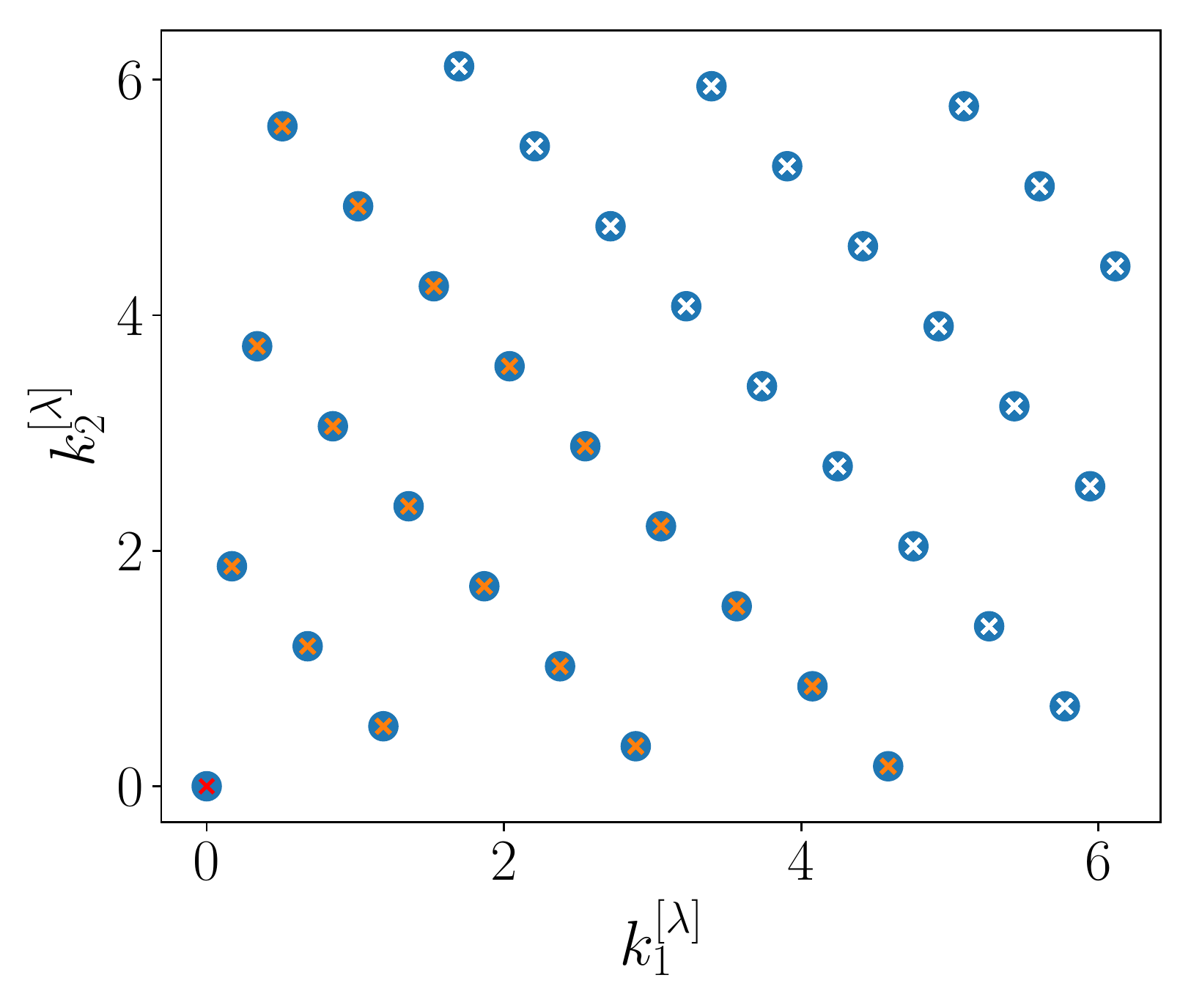}}
\hfil{}
\subfloat[]{\label{fig: rightks}\includegraphics[width = 0.42\textwidth]{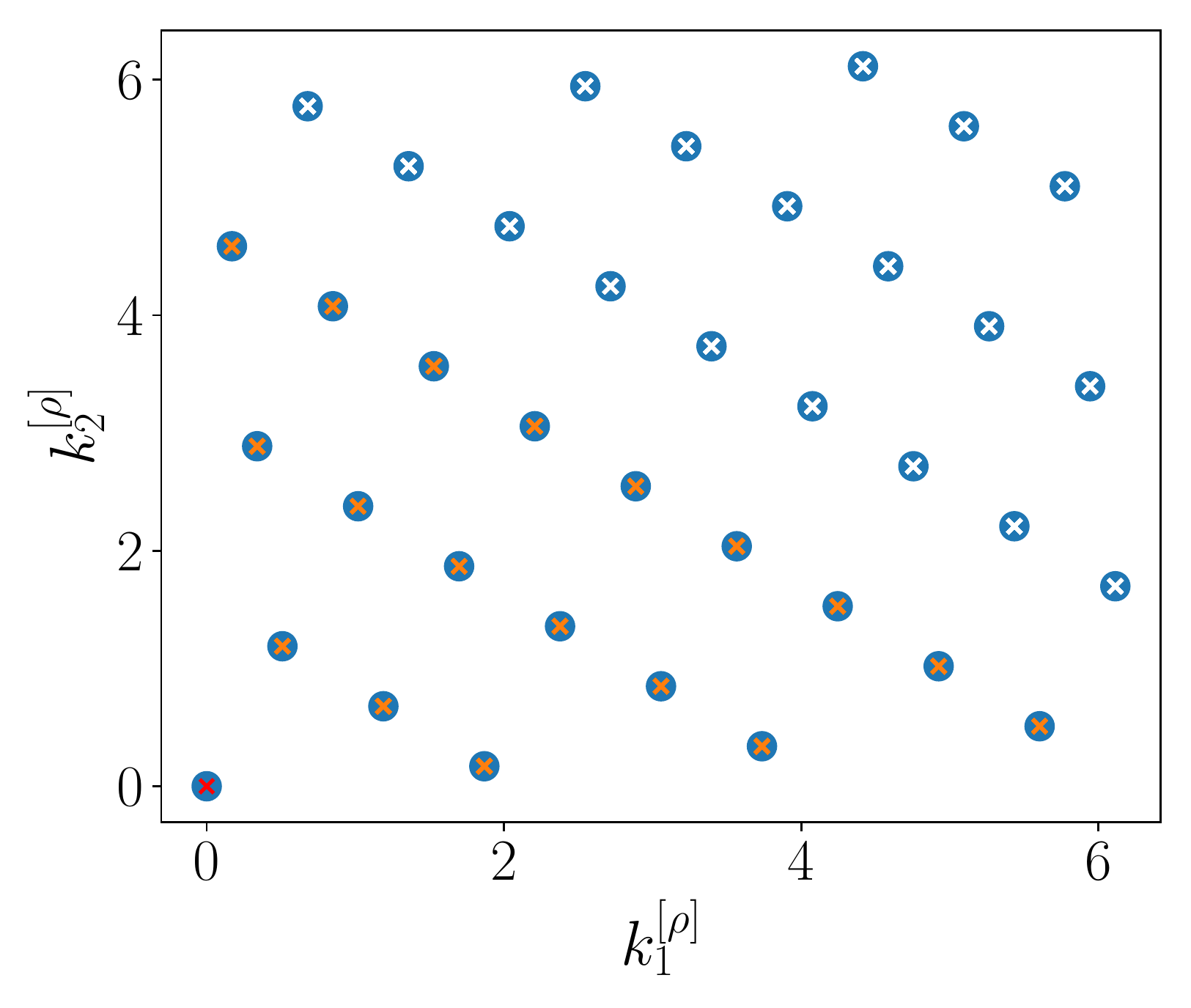}} \caption{The $k_1^{[i]}$,
$k_2^{[i]}$ value that give a singularity of the Dirac delta in Eq. (\ref{eq: bounded_prop3}) for an $R=3$ domain for
(a) the left shift and (b) the right shift. The large circles represent $k_1^{[i]}$, $k_2^{[i]}$ found numerically by
summing over $m_1$, $m_2$. The corresponding parameterised values, Eq. (\ref{eq: ks}) are represented by crosses. The
first three ($R$) diagonal rows correspond to the positive $k_1^{[i]}$, $k_2^{[i]}$ values, while the remaining three
are the negative ones, where for ease of visualisation, we have added a $2\pi$ phase to the negative terms. The term
corresponding to the steady state is shown at $(0,0)$. Note that in this case, there are $\Omega = 37$ points that need
to be paramaterised.}
\label{fig: ks}
\end{figure}

To proceed, one finds the values of $m_1$ and $m_2$, which lead to singularities that lie within the integral bounds
where each $(m_1, m_2)$ corresponds to a unique point in the finite domain. However, due to the non-orthogonality of the
coordinate points, one cannot independently sum $m_1$ and $m_2$ along the length of each axis, as one does for the
square lattice. To overcome this, we parameterise $k^{[i]}_1,k^{[i]}_2$ via Eq. (\ref{eq: ks}), alongside their
corresponding negative value, and create the nested summation in Eq. (\ref{eq: periodic-sol-hex-coords}) in the main
text. Figure \ref{fig: ks} shows the validity of this parameterisation for the left (panel (a)) and right (panel (b))
shift. Upon substituting these parameterised values into Eq. (\ref{eq: bounded_prop3}) and simplifying the
complex exponential, one obtains the exact periodic propagator Eq. (\ref{eq: periodic-sol-hex-coords}).

\section{Derivation of the Periodic Honeycomb Propagator}\label{honeycomb-app} Starting from Eq. (\ref{eq:
honeycomb_unbounded}) in the main text and applying the images defined in Sec. \ref{sec: Periodic} leads to 
\begin{equation}
\begin{aligned}
	\widetilde{\bm{\mathcal{P}}}^{(p)_{[i]}}_{\bm{n}_{0}, m_0}(n_1, n_2,  z) = \frac{1}{\Omega}\int_{-\pi}^{\pi}\int_{-\pi}^{\pi}\sum_{m_1 = -\infty}^{\infty}\sum_{m_2 = -\infty}^{\infty}  \delta(\bm{k} - 2\pi \mathbb{D}_{[i]}^{-1}\cdot\bm{m}^{\intercal})e^{i[(\bm{n} - \bm{n}_0)\cdot \bm{k}]}\left[\mathbb{I} -  z\bm{\mu}(k_1, k_2)\right]^{-1} \cdot \bm{U}_{m_0} dk_1 dk_2.
\end{aligned}
\end{equation}
Expanding $\left[\mathbb{I} -  z\bm{\mu}(k_1, k_2)\right]^{-1}$ in powers of $z$, $\left[\mathbb{I} - z\bm{\mu}(k_1,
k_2)\right]^{-1} = \sum_{t=0}^{\infty} \left[z\bm{\mu}(k_1, k_2)\right]^t$ and shifting the integral limits one has 
\begin{equation}
\begin{aligned}
\bm{\mathcal{P}}^{(p)_{[i]}}_{\bm{n}_{0}, m_0}(n_1, n_2, t) = \frac{1}{\Omega}\int_{-\varepsilon}^{2\pi-\varepsilon}\int_{-\varepsilon}^{2\pi-\varepsilon}\sum_{m_1 = -\infty}^{\infty}\sum_{m_2 = -\infty}^{\infty}\delta(\bm{k} - 2\pi \mathbb{D}_{[i]}^{-1}\cdot\bm{m}^{\intercal})e^{i[(\bm{n} - \bm{n}_0)\cdot \bm{k}]} \bm{\mu}(k_1, k_2)^t \cdot \bm{U}_{m_0} dk_1 dk_2.
\end{aligned}
\end{equation}
Using the known parametrisation of the Dirac delta (see Appendix \ref{app: periodic}), one finds the closed-form
solution for the honeycomb lattice, shown in Eq. (\ref{eq: honeycomb_periodic}) of the main text.

\subsection{Steady state behaviour in the honeycomb lattice}\label{sec: honey_ss} Since the eigenvalues and eigenvectors
of $\bm{\mu}\left(0, 0\right)$ are readily available, one can diagonalise the symmetric matrix $\bm{\mu}\left(0, 0\right) =
\mathbb{P}\mathbb{E}\mathbb{P}^{-1}$ where $\mathbb{E}_{1, 1} = 1$, $\mathbb{E}_{i,i} = 1-q$, $2 \leq i \leq 5$,
$\mathbb{E}_{6,6} = 1-2q$, $\mathbb{E}_{i,j} = 0$ otherwise, and 
\begin{equation}
    \mathbb{P} = \frac{1}{\sqrt{6}}\begin{bmatrix}
 -1 & 1 & 0 & -\sqrt{3} & 0 & -1 \\
 1 & 1 & -\sqrt{3} & 0 & -1 & 0 \\
-1 & 1 & 0 & 0 & 0 & 2 \\
 1 & 1 & 0 & 0 & 2 & 0 \\
-1 & 1 & 0 & \sqrt{3} & 0 & -1 \\
 1 & 1 & \sqrt{3} & 0 & -1 & 0 \\
\end{bmatrix}.
\end{equation}

For $0 < q < 1$, $\mathbb{P}\mathbb{E}^{t}\mathbb{P}^{-1} \to \mathbb{P}\overline{\mathbb{E}}\mathbb{P}^{-1}$ as $t \to
\infty$, where $\overline{\mathbb{E}}_{1,1} = 1$ and $\overline{\mathbb{E}}_{i,j} = 0$ otherwise.  Evaluating $
\mathbb{P}\overline{\mathbb{E}}\mathbb{P}^{-1}$, it is straightfoward to find $\bm{\mu}(0,0)^t = \frac{1}{6}\mathbb{J}$ as $t
\to \infty$.

For the $q=1$ case one has to be more careful as $ \lim_{t\to \infty}\mathbb{E}_{6,6} = \lim_{t\to \infty}(-1)^t$. Here,
$\bm{\mu}(0,0)$ is a hollow Toeplitz matrix with alternating bands of 0 and $\frac{1}{3}$. By inspecting this matrix, it
is clear that all odd powers of $t$ revert $\bm{\mu}(0,0)$ onto itself, while even powers of $t$ `swap' the bands,
giving rise to the alternating steady state probabilities in this case.

\section{Propagator with Absorbing Defects with Internal States}\label{sec: defect_internal_states} Here we outline the
generalisation of the defect technique on the lattice to random walks with internal states. It follows closely to, and
generalises, the derivation in Sec. 2.1 of \cite{luca_transmission}. 

We begin with the general Master equation governing the dynamics of the occupation probability of a random walker in a
lattice with defective internal states $(\bm{b}, m_{\bm{b}})$ in the set $\mathcal{B}$ with $\mathcal{N}$ defects
\begin{equation}
	\label{eq: internal_master_eq}
	\begin{aligned}
		\mathcal{P}(\bm{n}, m, t+1) &= \sum_{\bm{n'}}\sum_{m'}A(\bm{n}, m, \bm{n'}, m')\mathcal{P}(\bm{n'}, m', t), \; \; \; \bm{n}, m \notin \mathcal{B} \\
		\mathcal{P}(\bm{b}_i, m_{\bm{b}_i}, t+1) &= (1-\rho_{\bm{b}_i, m_{\bm{b}_i}}) \sum_{\bm{n'}}\sum_{m'}A(\bm{b}_i, m_{\bm{b}_i}, \bm{n'}, m')\mathcal{P}(\bm{n'}, m', t), \; \; \; \bm{b}_i, m_{\bm{b}_i} \in \mathcal{B}, 
	\end{aligned}
\end{equation} 
i.e. $i \in \{1, ..., |\mathcal{B}|\}$, $A(\bm{n}, m, \bm{n'}, m')$ is the transition probability tensor from state
$\bm{n}, m$ to state $\bm{n}', m'$, and where $\rho_{\bm{b}_i, m_{\bm{b}_i}}$ $(0 \leq \rho_{\bm{b}_i, m_{\bm{b}_i}}
\leq 1)$ governs the probability of getting absorbed at defect $\bm{b}_i, m_{\bm{b}_i}$ where $\rho_{\bm{b}_i,
m_{\bm{b}_i}} = 1$ represents perfect trapping efficiency at that site. To proceed, one first considers $\rho_{\bm{b}_i,
m_{\bm{b}_i}} \neq 1$ case. 

For convenience, we combine Eq. (\ref{eq: internal_master_eq}) into one equation 
\begin{equation}
\begin{aligned}
	\label{eq: internal_master_eq1}
	\mathcal{P}(\bm{n}, m, t+1) =  \sum_{\bm{n'}}\sum_{m'}\bigg[A(\bm{n}, m, \bm{n'}, m')\mathcal{P}(\bm{n'}, m', t) - \sideset{}
 {^\prime}\sum_{\bm{b}}\sum_{m_{\bm{b}}}\rho_{\bm{b}, m_{\bm{b}}} \delta_{\bm{n} \bm{b}}\delta_{m m_{\bm{b}}}A(\bm{b}, m_{\bm{b}}, \bm{n'}, m')\mathcal{P}(\bm{n}', m', t)\bigg],
	\end{aligned}
\end{equation}
where the primed summation is over all defective sites containing a defective state, and the following summation is over
all the states $m_{\bm{b}}$ in that site. The formal solution is simply the propagator of the defect free problem plus
the propagator convoluted in time and space with the known, defect-free term \cite{luca_transmission}. Calling
$\Psi_{\bm{n}_{0}, m_0}(\bm{n}, m, t)$ the defect-free propagator (the periodic propagator in this case) and applying
the localised initial condition $\mathcal{P}(\bm{n}, m, 0) = \delta_{\bm{n} \bm{n}_{0}} \delta_{m
m_0}\left[(1-\rho_{\bm{n}_{0} m_0})\delta_{\bm{n}_{0} m_0 \in \mathcal{B}} +\delta_{\bm{n}_{0} m_0 \notin \mathcal{B}}
\right]$ one obtains, in $z$-domain, 
\begin{equation}
\begin{aligned}
	\widetilde{\mathcal{P}}_{\bm{n}_{0}, m_0}(\bm{n}, m, z) = \widetilde{\Psi}_{\bm{n}_{0}, m_0}(\bm{n}, m, z) -  \sideset{}
 {^\prime}\sum_{\bm{b}}\sum_{m_{\bm{b}}} \frac{\rho_{\bm{b}, m_{\bm{b}}}}{1-\rho_{\bm{b}, m_{\bm{b}}}}\widetilde{\Psi}_{\bm{b}, m_{\bm{b}}}(\bm{n}, m, z)\widetilde{\mathcal{P}}_{\bm{n}_{0}, m_0}(\bm{b}, m_{\bm{b}}, z).
	\end{aligned}
	\label{eq: defect-to-solve}
\end{equation}
After setting $\bm{n}, m$ to all absorbing sites $\bm{b}, m_{\bm{b}}$, Eq. (\ref{eq: defect-to-solve}) can be solved via
Cramer's rule giving
\begin{equation}
	\label{eq: det_ratio}
	\widetilde{\mathcal{P}}_{\bm{n}_{0}, m_0}(\bm{b}_j, m_{\bm{b}_j}, z) = (1 - \rho_{\bm{b}_j, m_{\bm{b}_j}})\frac{\det(\mathbb{H}^{(j)}(\bm{\rho}, \bm{n}_{0}, m_0, z))}{\det(\mathbb{H}(\bm{\rho}, z))}. 
\end{equation}
Equation (\ref{eq: det_ratio}) represents the generating function of the probability of being at defective site
$\bm{b}_{j}$, $m_{\bm{b}_j}$ at time $t$ and not having been absorbed in any of the other sites in the set
$\mathcal{B}$. 

The elements in the matrix $\mathbb{H}(\bm{\rho}, z)$ are given as $\mathbb{H}_{k,k}(\bm{\rho}, z) = 1 +\rho_{\bm{b}_k,
m_{\bm{b}_k}} + \rho_{\bm{b}_k, m_{\bm{b}_k}} \widetilde{\Psi}_{\bm{b}_k, m_{\bm{b}_k}}(\bm{b}_k, m_{\bm{b}_k}, z)$ and
$\mathbb{H}_{i, k}(\bm{\rho}, z) = \rho_{\bm{b}_k, m_{\bm{b}_k}} \widetilde{\Psi}_{\bm{b}_k, m_{\bm{b}_k}}(\bm{b}_i,
m_{\bm{b}_i}, z)$ and $\mathbb{H}^{(j)}(\bm{\rho}, \bm{n}_{0}, m_0, z)$ is the same as $\mathbb{H}(\rho, z)$ but with
the $j^{\text{th}}$ column replaced with $\left(\widetilde{\Psi}_{\bm{n}_{0}, m_0}(\bm{b}_1, m_{\bm{b}_1}, z), ...
,\widetilde{\Psi}_{\bm{n}_{0}, m_0}(\bm{b}_{\mathcal{N}}, m_{\bm{b}_{\mathcal{N}}}, z)\right)^{\intercal}$. 

Substituting Eq. (\ref{eq: det_ratio}) into Eq. (\ref{eq: defect-to-solve}) and taking the limit $ \rho_{\bm{b}_i,
m_{\bm{b}_i}} \to 1$ gives us the defective propagator given in Eq. (\ref{honey_absorbing_propagator}) of the main text,
where we have taken $\rho_{\bm{b}_k, m_{\bm{b}_k}} = 1$, for all $k$, to model the fully absorbing boundary. As such, in Eq.
(\ref{honey_absorbing_propagator}), we have dropped the $\rho$ dependence in $\mathbb{H}$.

\section{Propagator with Inert Spatial Heterogeneities with Internal States}\label{app: defect_reflect_internal_states}
Here we outline the derivation of the defect technique to account for inert spatial heterogeneities in random walks with
internal states. It follows closely to and generalises \cite{sree_reflect} (see Section I of the Supplementary
Material). In this case, the defects appear in pairs $\mathcal{B} = \{\{(\bm{b}_1, m_{\bm{b}_1}), (\bm{b}'_1,
m_{\bm{b}'_1})\},...,\{(\bm{b}_{\mathcal{N}}, m_{\bm{b}_{\mathcal{N}}}), (\bm{b}'_{\mathcal{N}},
m_{\bm{b}'_{\mathcal{N}}})\}\}$, that is, in the case of the reflective propagator, the boundary states and their
respective neighbour across the periodic boundary.

The dynamics are given by 
\begin{equation}
\label{eq: reflect_is_me}
	\mathcal{P}(\bm{n}, m, t+1) = \sum_{\bm{n'}}\sum_{m'}A(\bm{n}, m, \bm{n'}, m')\mathcal{P}(\bm{n'}, m', t),
\end{equation}
when the walker is not on a defective site. Instead, when the walker is on any defective site we have 
\begin{equation}
\label{eq: defect_dyn1}
\begin{aligned}
 \mathcal{P}(\bm{b}_i, m_{\bm{b}_i}, t+1) = \sum_{\bm{n'}}\sum_{m'}A(\bm{b}_i, m_{\bm{b}_i}, \bm{n'}, m')\mathcal{P}(\bm{n'}, m', t) + 
\eta_{m_{\bm{b}'_i}, m_{\bm{b}_i}}\mathcal{P}(\bm{b}_i, m_{\bm{b}_i}, t) - \eta_{m_{\bm{b}_i}, m_{\bm{b}'_i}}\mathcal{P}(\bm{b}'_i, m_{\bm{b}'_i}, t) ,
\end{aligned}
\end{equation}
\begin{equation}
\label{eq: defect_dyn2}
\begin{aligned}
 \mathcal{P}(\bm{b}'_i, m_{\bm{b}'_i}, t+1) = \sum_{\bm{n'}}\sum_{m'}A(\bm{b}_i, m_{\bm{b}_i}, \bm{n'}, m')\mathcal{P}(\bm{n'}, m', t) + 
\eta_{m_{\bm{b}_i}, m_{\bm{b}'_i}}\mathcal{P}(\bm{b}'_i, m_{\bm{b}'_i}, t) - \eta_{m_{\bm{b}'_i}, m_{\bm{b}_i}}\mathcal{P}(\bm{b}_i, m_{\bm{b}_i}, t).
\end{aligned}
\end{equation}
Once again, combining Eqs. (\ref{eq: defect_dyn1}) and (\ref{eq: defect_dyn2}) into one equation and taking the
$z$-transform, one obtains
\begin{equation}
	\label{eq: reflect_is_2}
\begin{aligned}
	\widetilde{\mathcal{P}}(\bm{n}, m , z) - \mathcal{P}(\bm{n}, m , 0) = \sum_{\bm{n'}}\sum_{m'}A(\bm{n}, m, \bm{n'}, m')&\widetilde{\mathcal{P}}(\bm{n'}, m', z) + 
	z\sum_{i = 1}^{\mathcal{N}}\Big\{\delta_{\bm{b}_i \bm{n}}\delta_{m_{\bm{b}_i} m}  - \delta_{\bm{b}'_i \bm{n}}\delta_{m_{\bm{b}'_i} m} \Big\}\\ & \times\left[\eta_{m_{\bm{b}_i}, m_{\bm{b}'_i}}\widetilde{\mathcal{P}}(\bm{b}'_i, m_{\bm{b}'_i}, z) - \eta_{m_{\bm{b}'_i}, m_{\bm{b}_i}}\widetilde{\mathcal{P}}(\bm{b}_i, m_{\bm{b}_i}, z)\right],
\end{aligned}
\end{equation} 
with the parameters $\eta_{\bm{u}, \bm{v}}$ defined in the main text. Assuming the defect-free solution
$\widetilde{\Psi}_{\bm{n}_{0}, m_0}(\bm{n}, m, z)$ is known, i.e. the propagator of Eq. (\ref{eq: reflect_is_me}), which
in our case we again take as the periodic propagator, the general solution of Eq. (\ref{eq: reflect_is_2}) for a
localised initial condition $ \mathcal{P}(\bm{n}, m , 0) = \delta_{\bm{n} \bm{n}_0}\delta_{m m_0}$ is
\begin{equation}
\begin{aligned}
\widetilde{\mathcal{P}}_{\bm{n}_{0}, m_0}(\bm{n}, m, z) &= \widetilde{\Psi}_{\bm{n}_{0}, m_0}(\bm{n}, m, z)+  
z\sum_{j = 1}^{\mathcal{N}}\widetilde{\Psi}_{\langle \bm{b}_j,m_{\bm{b}_j}   - \bm{b}'_j,m_{\bm{b}'_j}\rangle}(\bm{n}, m, z) \\&
\times\left[\eta_{m_{\bm{b}_j}, m_{\bm{b}'_j}}\widetilde{\mathcal{P}}(\bm{b}'_j, m_{\bm{b}'_j}, z) - \eta_{m_{\bm{b}'_j}, m_{\bm{b}_j}}\widetilde{\mathcal{P}}(\bm{b}_i, m_{\bm{b}_i}, z)\right].
\end{aligned}
\end{equation}
We again solve by creating $\mathcal{N}$ simultaneous equations (for each defect pair). Using Cramer's rule we obtain
the propagator, 
\begin{equation}
\begin{aligned}
\widetilde{\mathcal{P}}_{\bm{n}_{0}, m_0}(\bm{n}, m, z) = \widetilde{\Psi}_{\bm{n}_{0}, m_0}(\bm{n}, m, z) -  \sum_{j = 1}^{\mathcal{N}}\widetilde{\Psi}_{\langle \bm{b}_j,m_{\bm{b}_j} - \bm{b}'_j,m_{\bm{b}'_j}\rangle}(\bm{n}, m, z) \frac{\det(\mathbb{S}(\bm{n}_{0}, m_0, z))}{\det(\mathbb{S}(z))},
\label{eq: ref_def1}
\end{aligned}
\end{equation}
where 
\begin{equation}
\begin{aligned}
\mathbb{S}(z)_{i,k} =\eta_{\bm{b}'_i, \bm{b}_i}\widetilde{\Psi}_{\langle \bm{b}_k,m_{\bm{b}_k}   -  \bm{b}'_k,m_{\bm{b}'_k}\rangle}(\bm{b}_i, m_{\bm{b}_i}, z) - \eta_{\bm{b}_i, \bm{b}'_i}\widetilde{\Psi}_{\langle \bm{b}_k,m_{\bm{b}_k}  - \bm{b}'_k, m_{\bm{b}'_k} \rangle}(\bm{b}'_i, m_{\bm{b}'_i}, z) -  \frac{\delta_{i k}}{z},
\end{aligned}
\end{equation}
and $\mathbb{S}(\bm{n}_{0}, m_0, z)$ the same as $\mathbb{S}(z)$ but with the $j^{\text{th}}$ column replaced with 
\begin{equation}
\begin{aligned}
&\Big[(\eta_{m_{\bm{b}'_1}, m_{\bm{b}_1}}\widetilde{\Psi}_{\bm{n}_{0}, m_0}(\bm{b}_1, m_{\bm{b}'_1}, z) - \eta_{m_{\bm{b}_1}, m_{\bm{b}'_1}} \widetilde{\Psi}_{\bm{n}_{0}, m_0}(\bm{b}'_1, m_{\bm{b}'_1}, z) , ..., \\ & \eta_{m_{\bm{b}'_{\mathcal{N}}}, m_{\bm{b}_\mathcal{N}}}\widetilde{\Psi}_{\bm{n}_{0}, m_0}( \bm{b}_{\mathcal{N}}, m_{\bm{b}_{\mathcal{N}}}, z) - \eta_{m_{\bm{b}_{\mathcal{N}}}, m_{\bm{b}'_\mathcal{N}}}\widetilde{\Psi}_{\bm{n}_{0}, m_0}( \bm{b}'_{\mathcal{N}}, m_{\bm{b}'_{\mathcal{N}}}, z) \Big]^{^\intercal}.
\end{aligned}
\end{equation}
Evaluating the sum in Eq. (\ref{eq: ref_def1}) explicitly and taking $\eta_{m_{\bm{b}_i}, m_{\bm{b}'_i}} = \eta_{m_{\bm{b}'_i}, m_{\bm{b}_i}} = \frac{q}{3}$, i.e. the outgoing probability from each site in the defect-free honeycomb lattice, one finds the honeycomb reflective
propagator as shown in Eq. (\ref{eq: reflect_honey_prop}) in the main text.
\section{Honeycomb MFPT and MRT}\label{app: honey_MFPT} 
\subsection{Periodic Boundary Conditions}\label{app: per_honey_MFPT}
Using the $z$-transform Eq. (\ref{eq: honeycomb_periodic}), one finds the return probability as
\begin{equation}
    \begin{aligned}
    &\mathcal{R}^{(p)_{[i]}}_{\bm{n}_0, m_0}(z) = 1 -\\& \frac{\Omega}{\bm{U}^{\intercal}_{m_0}\cdot\left[[\mathbb{I} - z\bm{\mu}(0,0)]^{-1} + \sum_{r=0}^{R-1}\sum_{s=0}^{3r+2} \Bigg\{\left[\mathbb{I} - z\bm{\mu}\left(\frac{2\pi k_1^{[i]}}{\Omega}, \frac{2\pi k_2^{[i]}}{\Omega}\right)\right]^{-1} +
    \left[\mathbb{I} - z\bm{\mu}\left(\frac{-2\pi k_1^{[i]}}{\Omega}, \frac{-2\pi k_2^{[i]}}{\Omega}\right)\right]^{-1} \Bigg\}\right]\cdot \bm{U}_{m_0}}.
     \label{eq: honey_return1}   
\end{aligned}   
\end{equation} 
Owing to the recurrence of the random walk, the matrix $[\mathbb{I} - z\bm{\mu}(0,0)]$ is singular at $z=1$. Denoting the summand
in Eq. (\ref{eq: honey_return1}) as $\mathbb{M}^{[i]}(r,s)$ and multiplying the top and bottom of the fraction by $\det([\mathbb{I} - z\bm{\mu}(0,0)])$, we find  
\begin{equation}
    \begin{aligned}
    \mathcal{R}^{(p)_{[i]}}_{\bm{n}_0, m_0}(z) = 1 - \frac{\Omega\det([\mathbb{I} - z\bm{\mu}(0,0)])}{\bm{U}^{\intercal}_{m_0}\cdot\left[\text{Inv}\left([\mathbb{I} - z\bm{\mu}(0,0)]\right) + \det([\mathbb{I} - z\bm{\mu}(0,0)])\sum_{r=0}^{R-1}\sum_{s=0}^{3r+2} \mathbb{M}^{[i]}(r,s)\right]\cdot \bm{U}_{m_0}},
     \label{eq: honey_return2}   
\end{aligned}   
\end{equation} 
where the notation $\text{Inv}(\cdot)$ denotes the inverse matrix multiplied by its determinant. Evaluating $\frac{\partial \mathcal{R}^{[i]}_{\bm{n}_0, m_0}}{\partial z}\Bigr|_{z=1}$, 
and utilising the property $\det([\mathbb{I} - z\bm{\mu}(0,0)])\Bigr|_{z=1}=0$, we obtain for either shift,
\begin{equation}
    \mathdutchcal{R}^{(p)}_{\bm{n}_0, m_0} = -\frac{\Omega\left (\frac{\partial \det([\mathbb{I} - z\bm{\mu}(0,0)])}{\partial z}\Bigr|_{z=1}\right )}{\bm{U}^{\intercal}_{m_0}\cdot\text{Inv}(\mathbb{I} - \bm{\mu}(0,0))\cdot \bm{U}_{m_0}}.
    \label{eq: honey_mrt}
\end{equation}
Upon inspection of the $6\times 6$ matrix, one finds $\frac{\partial \det([\mathbb{I} - z\bm{\mu}(0,0)])}{\partial z}\Bigr|_{z=1} = -2q^{5}$ and 
$\text{Inv}(\mathbb{I} - \bm{\mu}(0,0)) = \frac{q^5}{3}\mathbb{J}$, where $\mathbb{J}$ again denotes an all ones matrix. Using these values in Eq. (\ref{eq: honey_mrt}),
we confirm Kac's lemma obtaining $  \mathdutchcal{R}_{\bm{n}_0, m_0} =6\Omega$, the number of states in the domain.

Following the same procedure on the honeycomb first passage probability, one obtains Eq. (\ref{eq: honey_MFPT}) of the
main text. When $R=0$, the periodic honeycomb MFPT, Eq. (\ref{eq: honey_MFPT}), reduces to    
\begin{equation}
    \mathdutchcal{F}^{(p)}_{ m_0 \to m} = \sum_{i=1}^{6}\lambda_i\left(\bm{U}_m^{\intercal}\cdot\bm{u}_i\cdot \bm u_{i}^{\intercal}\cdot\bm{U}_{m_0} - \bm{U}_m^{\intercal}\cdot\bm{u}_i\cdot \bm u_{i}^{\intercal}\cdot\bm{U}_{m}\right),
    \label{eq: R0_MFPT}
\end{equation}
where $\bm{u}_i$ are an orthonormal basis formed of the right eigenvectors associated with the eigenvalues of
$\mathbb{C}$, $\lambda_{k} =  \sum_{l=0}^{5}c_l e^{\frac{2\pi i(k-1)l}{6}} $, $k \in \{1, 6\}$. Evaluating Eq. (\ref{eq: R0_MFPT}), one finds
\begin{equation}
    \mathdutchcal{F}^{(p)}_{ m_0 \to m} = \frac{5}{q}\left(\delta_{|m-m_0|,1} + \delta_{|m-m_0|,3} +\delta_{|m-m_0|,5} \right) + \frac{6}{q}\left(\delta_{|m-m_0|,2} + \delta_{|m-m_0|,4} \right).
    \label{eq: R0_MFPT1}
\end{equation}
As expected due to the periodicity of the internal states, Eq. (\ref{eq: R0_MFPT1}) is symmetric
around $m_0, m$ and equals zero when $m_0 = m$.
\subsection{Reflective Boundary Conditions}\label{app: ref_honey_MFPT}
In the reflective honeycomb domain, implementing Appendix II of \cite{sree_reflect} to random walks with internal states, one 
obtains
\begin{equation}
	\label{eq: honey_reflect_MFPT}
	\mathdutchcal{F}^{(r)}_{\bm{n}_{0}, m_0 \to \bm{n}, m} = \mathdutchcal{F}^{(p)}_{\bm{n}_{0}, m_0 \to \bm{n}, m} - 1 + \frac{\det(\mathbb{F}-\mathbb{F}^{(1)})}{\det(\mathbb{F})},
\end{equation}
where 
\begin{equation}
	\label{eq: honey_MFPT_mat1}
	\mathbb{F}_{ij} = \frac{q}{18\Omega}\left[\mathdutchcal{F}^{(p)}_{\langle \bm{b}_j,m_{\bm{b}_j}   - \bm{b}'_j,m_{\bm{b}'_j} \rangle \to \bm{b}_i, m_{\bm{b}_i} } - \mathdutchcal{F}^{(p)}_{\langle \bm{b}_j,m_{\bm{b}_j}   - \bm{b}'_j,m_{\bm{b}'_j} \rangle \to \bm{b}'_i, m_{\bm{b}'_i} }\right] + \delta_{ij},
\end{equation}
and 
\begin{equation}
\label{eq: honey_MFPT_mat2}
	\mathbb{F}^{(1)}_{ij} = \frac{q\mathdutchcal{F}^{(p)}_{\langle \bm{b}_j,m_{\bm{b}_j}   - \bm{b}'_j,m_{\bm{b}'_j} \rangle \to \bm{n}, m}}{18\Omega}\left[\mathdutchcal{F}^{(p)}_{\langle \bm{n}_{0}, m_0 - \bm{n}, m \rangle \to \bm{b}_i, m_{\bm{b}} } - \mathdutchcal{F}^{(p)}_{\langle \bm{n}_{0} - \bm{n} \rangle \to \bm{b}'_i, m_{\bm{b}'_i} }\right],
\end{equation}
for the MFPT, while for the MRT we have $\mathdutchcal{R}^{(r)}_{\bm{n}, m} =
\mathdutchcal{R}^{(p)}_{\bm{n}, m} $, as expected. The factor $\frac{q}{18\Omega}$ in Eqs. (\ref{eq:
honey_MFPT_mat1}) and (\ref{eq: honey_MFPT_mat2}) is obtained via the simple multiplication of the periodic return
probability and the probability of movement, $\frac{q}{3}$.

\section{Efficiency of Computational Procedures}
To obtain propagators in the absorbing and reflective cases, e.g. evaluating Eqs. 
(\ref{hex_absorbing_propagator})-(\ref{eq: reflect_honey_prop}), or to obtain the splitting probabilities in
Eq. (\ref{eq: splitting}), one undertakes the numerical inverse $z$-transform \cite{z-inverse}, which
consists of evaluating
\begin{equation}
    f(t) = \frac{1}{2\pi i} \oint\limits_{|z| < 1} \frac{\widetilde{f}(z)}{z^{t+1}} dz\simeq \frac{1}{t r^t}\sum_{k=1}^{t-1}(-1)^k \mbox{Re}\left[\widetilde{f}\left( re^{\frac{ik\pi}{t}}\right)\right]+\frac{1}{2t r^t}\left[\widetilde{f}\left( r\right)+(-1)^t\widetilde{f}\left(-r\right)\right],
    \label{eq: z_inverse}
\end{equation} 
which has an error $e_r$ given by $e_r\leq r^{2t}/\left(1-r^{2t}\right)^{-1}$. The computational cost of this procedure
scales as a function of the number of lattice sites in the domain multiplied by the number of defective sites squared,
multiplied by the time $t$. To illustrate let us consider the hexagonal absorbing propagator. The nested double
summation to obtain the periodic propagator scales with the size of the domain, $\Omega \sim R^2 $. With $6R$ defects on
the boundary, the time complexity of the numerical inverse scheme scales quartically in $R$, i.e. $36R^2\Omega t\sim R^4
t$. 

The computation required for the honeycomb lattice is very similar but with the slight additional burden of computing
the inverse of the $6\times 6$ matrix in Eq. (\ref{eq: honeycomb_unbounded}), which increases the computation time by a
scale factor of $c$ ($6^{2.37} \lesssim c \leq 6^3$) depending on which numerical scheme is used
\cite{williams2012multiplying}. 

To obtain the first passage to multiple targets, one must populate the matrix in Eq. (\ref{eq: splitting}) creating a
computational cost that scales as $ E^2 R^4 t$, where $E$ is the number of targets. This scaling should be compared to
an alternate procedure to calculate the FP to multiple sites, which consists of iteratively solving a Master equation
\cite{PhysRevLett.95.260601}. Convenient implementation of such procedure in hexagonal geometry would require utilising
the relationship between HCC and $3$-dimensional Cartesian coordinates i.e. Eq. (\ref{eq: original-master-eq}) modified
appropriately to account for the chosen boundary conditions. One would then be required to iteratively solve a
sixth-order sparse tensor and extract information from $E$ targets at each time iteration and then set those values to
zero, which would scale as $E R^{6} t$. 

A further advantage of our approach comes when extracting the means of random walk statistics. To obtain MFPTs, by setting $z=1$, one
bypasses the need to compute an inverse $z$-transform. As such, for the hexagonal lattice, the complexity for the
reflective MFPT scales as $ E^2 R^4$, where $E$ is the number of targets. If, on the other hand, one were to compute this via an
iterative method, an entire transmission probability would need to be obtained, which is far more computationally
expensive and introduces the uncertainty of a stopping criterion to approximate long time indirect trajectories. 

Note that performing stochastic simulations instead of the analytic techniques developed is also disadvantageous. The main
reason stems from the impossibility to  reduce systematically the error in the simulation output as one increases the
size of the ensemble. One is then forced to run a large, time expensive, ensemble, which limits the ability to explore
the parameter space.
\bibliography{references}
\bibliographystyle{unsrt}
\end{document}